%% file: upper_bound_transactions_camera_ready_two_column_version5.tex
\documentclass[10pt]{IEEEtran}

\usepackage[utf8]{inputenc} 
\usepackage[T1]{fontenc}
\usepackage{url}
\usepackage{ifthen}
\usepackage{cite}
\usepackage[cmex10]{amsmath} % Use the [cmex10] option to ensure complicance
\usepackage{amsfonts,amssymb,dsfont}
\usepackage{graphicx}
\usepackage[]{algorithm2e}

\usepackage{epstopdf}
\usepackage{enumerate}

\usepackage{tikz}
\usetikzlibrary{positioning,chains,fit,shapes,calc,arrows}
\usepackage{verbatim,booktabs}
\usepackage{notation}

\usepackage{pgfplots}
\usetikzlibrary{plotmarks}

\newtheorem{theorem}{Theorem}
\newtheorem{proposition}{Proposition}

\newtheorem{lemma}{Lemma}
\newtheorem{example}{Example}
\newtheorem{definition}{Definition}

\newtheorem{remark}{Remark}
\DeclareMathOperator*{\argmax}{arg\,max}
\DeclareMathOperator*{\argmin}{arg\,min}

\ifodd 1 \newcommand{\rev}[1]{{\color{blue}#1}}
\else \newcommand{\rev}[1]{#1} \fi

\ifodd 1 
\else  \fi

\IEEEoverridecommandlockouts

\begin{document}
	\title{A Single-Letter Upper Bound \\to the Mismatch Capacity} 
	
	% %%% Single author, or several authors with same affiliation:
	% \author{%
	%   \IEEEauthorblockN{Stefan M.~Moser}
	%   \IEEEauthorblockA{ETH Zürich\\
	%                     ISI (D-ITET)\\
	%                     CH-8092 Zürich, Switzerland\\
	%                     Email: moser@isi.ee.ethz.ch}
	% }

	\author{Ehsan Asadi Kangarshahi and Albert Guill\'en i F\`abregas
		\thanks{E. Asadi Kangarshahi is with the Department of Engineering, 
			University
			of Cambridge, Cambridge CB2 1PZ, U.K. (e-mail: ea460@cam.ac.uk).
			
			A.~Guill\'en i F\`abregas is with the Department of Information and 
			Communication
			Technologies, Universitat Pompeu Fabra, Barcelona 08018, Spain,
			also with the Instituci\'o Catalana de Recerca i Estudis Avan\c{c}ats 
			(ICREA),
			Barcelona 08010, Spain, and also with the Department of Engineering, 
			University
			of Cambridge, Cambridge CB2 1PZ, U.K. (e-mail: guillen@ieee.org).
			
			This work was supported in part by the European Research Council under 
			Grant 725411, and by the Spanish Ministry of Economy and Competitiveness 
			under Grant TEC2016-78434-C3-1-R.
		}
		\thanks{This work has been presented in part at the 2019 IEEE International Symposium on Information Theory, Paris, France, and at the 2020 International Z\"urich Seminar on Information and Communication.}
	}

	%%% Many authors with many affiliations:
	% \author{%
	%   \IEEEauthorblockN{Albus Dumbledore\IEEEauthorrefmark{1},
	%                     Olympe Maxime\IEEEauthorrefmark{2},
	%                     Stefan M.~Moser\IEEEauthorrefmark{3}\IEEEauthorrefmark{4},
	%                     and Harry Potter\IEEEauthorrefmark{1}}
	%   \IEEEauthorblockA{\IEEEauthorrefmark{1}%
	%                     Hogwarts School of Witchcraft and Wizardry,
	%                     1714 Hogsmeade, Scotland,
	%                     \{dumbledore, potter\}@hogwarts.edu}
	%   \IEEEauthorblockA{\IEEEauthorrefmark{2}%
	%                     Beauxbatons Academy of Magic,
	%                     1290 Pyrénées, France,
	%                     maxime@beauxbatons.edu}
	%   \IEEEauthorblockA{\IEEEauthorrefmark{3}%
	%                     ETH Zürich, ISI (D-ITET), ETH Zentrum, 
	%                     CH-8092 Zürich, Switzerland,
	%                     moser@isi.ee.ethz.ch}
	%   \IEEEauthorblockA{\IEEEauthorrefmark{4}%
	%                     National Chiao Tung University (NCTU), 
	%                     Hsinchu, Taiwan,
	%                     moser@isi.ee.ethz.ch}
	% }
	
	%\date{\today}
	
	\maketitle
	
	%%%%%%
	%% Abstract: 
	%% If your paper is eligible for the student paper award, please add
	%% the comment "THIS PAPER IS ELIGIBLE FOR THE STUDENT PAPER
	%% AWARD." as a first line in the abstract. 
	%% For the final version of the accepted paper, please do not forget
	%% to remove this comment!
	%%
	\begin{abstract}
		We derive a single-letter upper bound to the mismatched-decoding capacity for discrete memoryless channels. The bound is expressed as the mutual information of a transformation of the channel, such that a maximum-likelihood decoding error on the translated channel implies a mismatched-decoding error in the original channel. In particular, it is shown that if the rate exceeds the upper-bound, the probability of error tends to one exponentially when the block-length tends to infinity. We also show that the underlying optimization problem is a convex-concave problem and that an efficient iterative algorithm converges to the optimal solution. In addition, we show that, unlike achievable rates in the literature, the multiletter version of the bound cannot not improve. A number of examples are discussed throughout the paper.

	\end{abstract}

	%% The paper must be self-contained. However, if you are referring to
	%% a full version for checking certain proofs, please provide the
	%% publically accessible location below.  If the paper is completely
	%% self-contained, you can remove the following line from your
	%% submission.
	%\textit{A full version of this paper is accessible at:}
	%\url{http://isit2019.fr/} 
	
	%%%%%%%%%%%%%%%%%%%%%%%%%%%%%%%%%%%%%%%%%%%%%
	
	\section{Introduction and Preliminaries}
	
	We consider reliable communication over a discrete memoryless channel (DMC) $W$ with a given decoding metric \cite{csiszarDMC,Merhav,huimis,csiszargraph} (see also \cite{scarlett2020fnt} and references therein for an account of recent progress). This problem arises when the decoder uses a suboptimal decoding rule due to limited computational resources, simpler implementation, lack of awareness of the channel law or imperfect channel estimation. Moreover, it is shown in \cite{csiszarDMC} that some important problems in information theory, like the zero-error capacity of a channel can be cast as instances of the mismatch decoding problem. As a result, deriving a single letter characterization of the mismatch decoding capacity would yield a solution to zero-error capacity problem, known to be a difficult problem. 
	
	Multiple achievability results have been reported in the literature \cite{huimis,csiszarDMC,csiszargraph,Merhav,lapidoth,somekh2014achievable,scarlett2016multiuser}. These results were derived by random coding techniques, \text{i.e.} analyzing the average probability of error of mismatch decoding over a certain ensemble of randomly generated codebooks. In some cases, multiuser achievable rates have been shown to improve over standard single-user random coding \cite{lapidoth,somekh2014achievable,scarlett2016multiuser}. As suggested by \cite{csiszarDMC}, multiletter versions of achievable rates can yield strict improvements over their single-letter counterparts.%Cost constrained random coding arguments were used in \cite{ganti,scarlett2014mismatched,scarlett2014reliable} to improve the code ensemble.
	
	Unlike the achievable rate case, few converse results have been reported in the literature. The only single-letter converse was reported in \cite{balakirsky1995converse}, where it was claimed that for binary-input DMCs, the mismatch capacity was precisely equal to the achievability result derived in \cite{huimis,csiszargraph} known as the LM rate. Reference \cite{counterexample} provided a counterexample to this converse invalidating its claim, showing that a multiletter multiuser rate from \cite{somekh2014achievable,scarlett2016multiuser} was strictly higher than the LM rate. Multiletter converse results were derived in \cite{aneliaconvers}. In particular,  for DMCs, \cite{aneliaconvers} shows that for rational decoding metrics, the probability of error cannot decay faster than $O(n^{-1})$ for rates above the achievable rate in \cite{huimis,csiszargraph}.
	
	In this paper, we propose a single-letter upper bound to the mismatch capacity that is shown to characterize the mismatch capacity in special cases where it is known, and yield strict improvements over the matched capacity in cases where the mismatch capacity is unknown. The bound is expressed as the mutual information of an auxiliary channel, such that a maximum-likelihood decoding error on the auxiliary channel implies a mismatched-decoding error in the original channel. The key is to connect the real and auxiliary channels by means of a graph in the output space. This is a new technique to derive upper bounds that could also be helpful in other settings.
	%The bound is cast as a max-min optimization of the mutual information between the input and a transformation of the channel, such that a maximum-likelihood decoding error on the transformed channel implies a mismatched-decoding error in the original channel. 
	The bound is shown to be convex-concave and an efficient algorithm to compute the bound is provided. The convexity analysis of the bound shows that the multiletter version cannot improve over its single-letter version.
	
	The paper is structured as follows. In Section \ref{sec:notation} we introduce notation and preliminaries. In Section \ref{smainresult} we introduce our main result and discuss its application to some examples.  Sections \ref{graph}, \ref{maximal},  \ref{continuous} and \ref{sec:end} provide the proof of our main result. In particular, in Section \ref{graph}, we construct a graph between different conditional type classes as a key first step of the proof of our upper bound. In Section \ref{maximal}, we relate the maximum-likelihood decoding errors on a constructed auxiliary channel $V$ and  mismatched decoding errors on channel $W$. In Section \ref{continuous} we extend the validity of the results derived in the previous sections, originally derived for types, to distributions. Section \ref{sec:end} gives the final steps of the proof. In Section \ref{convexity} we show that the optimization problem implied by our bound is a convex-concave optimization problem and we derive the corresponding KKT conditions. Section \ref{sec:comp} discusses the computation of the bound and proves the convergence of an efficient iterative algorithm based on the mirror prox algorithm \cite{Nemirovski}. In Section \ref{multilettersec} we use the KKT conditions derived for the single-letter bound and show that the multiletter version of the bound cannot improve over its single-letter counterpart. 
	
	%%%%%%%%%%%%%%%%%%%%%%%%%%%%%%%%%%%%%%%%%%%%%
	\section{Notation and Preliminaries}
	\label{sec:notation}
	We assume input and output alphabets are $\Xc = \{1,2,\dotsc,J\}$ and $\Yc = \{1,2,\dotsc,K\}$, respectively. We denote the channel transition probability by $W(k|j)$ and define $\Wm\in\RR^{J \times K}$ as the matrix defined by the channel $\Wm(j,k) = W(k|j)$. A codebook $\Cc_n$ is defined as a set of $M$ sequences $\Cc_n = \big\{\xv(1),\xv(2),\dotsc,\xv(M)\big\}$, where $\xv(m)= \big(x_1(m),x_2(m),\dotsc,x_n(m)\big)\in\Xc^n$, for $m \in \{1,2,\dotsc,M\}$. 
	A message $m \in \{1,2,\dotsc,M\}$ is  chosen equiprobably and $\xv(m)$ is sent over the channel. The channel produces a noisy observation $\yv=(y_1,y_2,\dotsc,y_n)\in\Yc^n$ according to $W^n(\yv|\xv) = \prod_{i = 1}^{n}W(y_i|x_i)$. 
	Upon observing $\yv\in\Yc^n$ the decoder produces an estimate of the transmitted message $\hat m \in \{1,2,\dotsc,M\}$. The average and maximal error probabilities are respectively defined as 
	\beq
	P_e(\Cc_n) = \frac 1 M \sum_{i=1}^M\PP[\hat m\neq m|m=i]
	\eeq
	and 
	\beq
	P_{e,\rm max}(\Cc_n) = \max_{i \in \{1,2,\dotsc,M\}}\PP[\hat m\neq m|m=i].
	\eeq 
	Rate $R>0$ is said to be achievable if for any $\epsilon > 0$ there exists a sequence of length-$n$ codebooks $\{\Cc_n\}_{n = 1}^{\infty}$ such that $|\Cc_n| \geq 2^{n(R-\epsilon)}$, and $ \liminf_{n \to \infty}P_e(\Cc_n)= 0$. The capacity of $W$, denoted by $C(W)$ or $C(\Wm)$, is defined as the largest achievable rate.
	
	The decoder that minimizes the error probability is the maximum-likelihood (ML) decoder, that produces the message estimate $\hat m$ according to
	\beq
	\hat m = \argmax_{i\in\{1,2,\dotsc,M\}} W^n\big(\yv|\xv(i)\big).
	\eeq
	In certain situations, where the decoder is unaware of the channel law, or is unable to compute it, it is not possible to use ML decoding and instead, the decoder produces the message estimate $\hat m$ as
	\beq
	\hat m = \argmax_{i\in\{1,2,\dotsc,M\}} \metric\big(\xv(i),\yv\big),
	\eeq
	where,
	\begin{align}\label{pvigtufunee}
		\metric\big(\xv(i),\yv\big) =\sum_{\ell=1}^n \metric\big(x_\ell(i), y_\ell\big)
	\end{align}
	and $\metric:\Xc\times\Yc\to\RR$ is the decoding metric\footnote{In the literature, $\metric(x,y)$ has been used to denote either an additive or multiplicative decoding metric. For convenience, we have chosen it to denote an additive metric.}. We assume that, without loss of generality, decoding ties are counted as errors. We will refer to this decoder as $\metric$-decoder. 
	When $\metric(x,y)=\log W(y|x)$, the decoder is ML, otherwise, for a general decoding metric $\metric$ the decoder is said to be mismatched \cite{huimis,csiszarDMC,csiszargraph,Merhav}. We define the metric matrix $\Qm\in \RR^{J \times K}$ with entries $\Qm(j,k) = \metric(j,k)$. The average and maximal error probabilities of codebook $\Cc_n$ under $\metric$-decoding are respectively denoted by $P_e^\metric(\Cc_n)$ and  $P_{e,\rm max}^\metric(\Cc_n)$.  The mismatch capacity $C_\metric(W)$ or $C_\metric(\Wm)$ is defined as supremum of all achievable rates with $\metric$-decoding. 
	
	Lower bounds for the mismatch capacity have been studied extensively using random coding techniques. Specifically, the i.i.d. random coding ensemble is known to achieve the generalized mutual information (GMI) which can be written as \cite{kaplan1993information},
	\begin{align}\label{gmirate}
		R^{\textsc{gmi}}_\metric(W) = \max_{P_X} \min_{\substack{V: \\ \EE_{P_X \times V}[\metric(X,Y)] \geq \EE_{P_X \times W}[\metric(X,Y)] }} I(P_X,V),
	\end{align}
	where the notation $P_X \times P_{Y|X}$ denotes the joint distribution induced by the corresponding marginal and conditional distributions.
	An improved lower bound, known as the LM rate, is derived by employing constant composition random coding \cite{csiszargraph,huimis},
	\begin{align} \label{lmrate}
		R^{\textsc{lm}}_\metric(W) = \max_{P_X} \min_{\substack{V: \\ P_X V = P_X W  \\ \EE_{P_X \times V}[\metric(X,Y)] \geq \EE_{P_X \times W}[\metric(X,Y)]}} I(P_X,V),
	\end{align}
	where the notation $P_X P_{Y|X}$ denotes the output distribution induced by the marginal distribution $P_X$ and conditional distribution $P_{Y|X}$. 
	The above rate has an intuitive explanation. The maximization is over all input distributions, and the minimizations is over all auxiliary channels $V$ with two properties. First, equal output marginal $P_X V = P_X  W$, such that for all $k \in \Yc$
	\begin{align}
		\sum_{j \in \Xc}P_X(j) V(k|j) = \sum_{j \in \Xc}P_X(j)W(k|j).
	\end{align}
	This implies that the distribution of the received sequence needs to be the same for both channel $W$ and auxiliary channel $V$ whenever the input codeword is chosen from composition $P_X$. The second condition, also present in the expression of the GMI, $\EE_{P_X \times V}[\metric(X,Y)] \geq \EE_{P_X \times W}[\metric(X,Y)] $ can be rewritten as,
	\begin{align}
		\sum_{j,k}^{}P_X(j)V(k|j)\metric(j,k) \geq \sum_{j,k}^{}P_X(j)W(k|j)\metric(j,k),
	\end{align}
	and implies that, the received sequence $Y$ has a higher metric under channel $V$ than under channel $W$, and therefore,
	%typical with one of the codewords under channel $V$, 
	the $\metric$-decoder makes an error. It is implied in \eqref{gmirate} and \eqref{lmrate} that $R^{\rm GMI}_\metric(W)\leq R^{\rm LM}_\metric(W)$. The GMI and LM rates are ensemble tight, \textit{i.e.} the ensemble average error probability tends to one exponentially for rates exceeding the GMI and LM rates, respectively. Both of the bounds above are known not to attain the mismatch capacity in general. 
	It is known that the GMI and LM rates can be improved by considering their multiletter counterparts \cite{csiszarDMC}.

	%%%%%%%%%%%%%%%%%%%%%%%%%%%%%%%%%%%%%%%%%%%%%
	%\section{Notation} \label{notation}
	
	The method of types \cite[Ch. 2]{csiszar2011information} will be used extensively in this paper. We recall some of the basic definitions and introduce some notation. The type of a sequence $\xv = (x_1,x_2,\dotsc,x_n) \in \Xc^n$ is the empirical distribution of its symbols, \textit{i.e.}, $ \hat{\pv}_{\xv} (j) = \frac{1}{n} \sum_{i = 1}^{n}\indicator\{x_i = j\}$. The set of all types of $\Xc^n$ is denoted by $\Pc_n(\Xc)$. For $\pv_X \in \Pc_n(\Xc)$, the type class $\Tc^n(\pv_X)$ is set of all sequences in $\Xc^n$ with type $\pv_X$, $\Tc^n(\pv_X) = \{\xv \in \Xc^n\,|\, \hat{\pv}_{\xv} = \pv_X\}$. 
	
	The joint type of sequences $\xv = (x_1,x_2,\dotsc,x_n)\in \Xc^n$ and $\yv=(y_1,y_2,\dotsc,y_n)  \in \Yc^n$ is the empirical distribution $\hat{\pv}_{\xv\yv}(j,k) = \frac{1}{n}\sum_{i = 1}^{n}\indicator\{x_i = j,y_i=k\}$. The conditional type of $\yv$ given $\xv$ is the empirical conditional distribution
	\begin{align}
		\hat{\pv}_{\yv|\xv}(k|j) = \begin{cases} 
			\frac{\hat{\pv}_{\xv\yv}(j,k)}{\hat{\pv}_{\xv}(j)} \
			&\hat{\pv}_{\xv}(j)> 0\\
			\frac{1}{K} \ &\text{otherwise.}
		\end{cases}
	\end{align}
	
	The set of all conditional types on $\Yc^n$ given $\Xc^n$ is denoted by $\Pc_n(\Yc|\Xc)$. For $\pv_{Y|X} \in \Pc_n(\Yc|\Xc)$ and a sequence $\xv \in \Tc^n(\pv_X)$, the conditional type class $\Tc^n_{\xv}(\pv_{Y|X})$ is defined as $\Tc^n_{\xv}(\pv_{Y|X}) = \{\yv \in \Yc^n\,|\, \hat{\pv}_{\yv|\xv} = \pv_{Y|X}\}.$
	
	Similarly, we can define the joint type of $\xv,\yv,\hat{\yv}$, as the empirical distribution of the triplet. For $j \in \Xc$ and $k_1,k_2 \in \Yc$,
	\begin{align}
		\hat{\pv}_{\xv\yv\hat{\yv}}(j,k_1,k_2) = \frac{1}{n}\sum_{i = 1}^{n} \indicator\{x_i = j,y_i = k_1,\hat{y}_i = k_2\}.
	\end{align}
	We define the joint conditional type of $\yv,\hat{\yv}$ given $\xv\in\Tc^n(\pv_X)$ as
	
	\begin{align}\label{definetype}
		\hat{\pv}_{\yv\hat{\yv}|\xv}(k_1,k_2|j) = \begin{cases} 
			\frac{\hat{\pv}_{\xv\yv\hat{\yv}}(k_1,k_2|j)}{\hat{\pv}_{\xv}(j)} \
			&\hat{\pv}_{\xv}(j)> 0\\
			\frac{1}{K}\indicator\{k_1=k_2\} \ &\text{otherwise.}
		\end{cases}
	\end{align}
	
	The set of all joint conditional types is denoted by $\Pc_n(\Yc \times \hat{\Yc}|\Xc)$. Additionally, for $\pv_{Y\hat{Y}|X} \in \Pc_n(\Yc\times\hat{\Yc}|\Xc)$ we define: 
	\begin{align}\label{lfnrnfgrn}
		\Tc^n_{\yv\xv}(\pv_{Y\hat{Y}|X}) = \{\hat{\yv} \in \Yc^n \,|\, \hat{\pv}_{\yv\hat{\yv}|\xv} = \pv_{Y\hat{Y}|X}\}.
	\end{align}
	The mutual information is defined as $I(P_X,P_{Y|X}) \eqdef \EE \Big[\log\frac{P_{Y|X}(Y|X)}{\sum_{x'} P_X(x') P_{Y|X}(Y|x')}\Big]$.
	Throughout the paper, for conditional types or conditional distributions $\Mm_1,\Mm_2$  we define
	\begin{align}
		|\Mm_1-\Mm_2|_{\infty} = \max_{\substack{1\leq j \leq J\\ 1 \leq k \leq K}}\big|\Mm_1(k|j)-\Mm_2(k|j)\big|.
	\end{align}
	
	\begin{definition} \label{fewer4tqw}
		Let $P_{Y\hat{Y}|X}$ be a joint conditional distribution and define the set
		\beq
		\Sc_\metric(k_1,k_2)\eqdef\big \{ i \in \Xc | i=\argmax_{i'\in\Xc}\metric(i',k_2) - \metric(i',k_1)\big\}.  
		\eeq
		We say that   $P_{Y\hat{Y}|X}$ is a maximal joint conditional distribution if for all $(j,k_1,k_2) \in \Xc \times \Yc \times \Yc$,
		\begin{align} \label{grre}
			P_{Y\hat{Y}|X}(k_1,k_2|j) = 0 ~\text{ if } ~j \notin \Sc_\metric(k_1,k_2).
		\end{align}
		Moreover, if $\pv_{Y\hat{Y}|X} \in \Pc_n(\Yc\times\hat{\Yc}|\Xc)$ satisfies the same condition, we call it a maximal joint conditional type. 
		
		For a given decoding metric $\metric$, we define the set of maximal joint conditional distributions to be $\Mc_{\rm max}(\metric)$. 
	\end{definition}
	
	Appendix C discusses the above definition for cases where the decoding metric $\metric$ can take $-\infty$ values.
	
	The above definition will become helpful when relating decoding errors in channel $P_{Y|X} = W$ under $\metric$-decoding to errors in channel $P_{\hat Y|X}$ under ML decoding.

	% \begin{definition}
	% Let $\Cc_n$ be a codebook. We say $\hat{\Cc_n}$ is a sub-codebook of $\Cc_n$ if $\hat{\Cc}_n \subset \Cc_n$.
	% \end{definition}
	
	\begin{definition} \label{bodsjgawdfabwd}
		Let $\Cc_n = \{\xv(1),\dotsc,\xv(M)\}$ and $m$ be the transmitted message. We say that the decoder makes a {\em type conflict error} for a given $\yv \in \Yc^n$ if there is at least one codeword $\xv(i)\neq \xv(m)$ such that $\hat{\pv}_{\yv|\xv(i)} = \hat{\pv}_{\yv|\xv(m)}$.
	\end{definition}
	
	If there is a type conflict error, every decoder that makes a decision based on the joint type between the channel output and the candidate codewords ($\alpha$-decoder) makes an error, including ML and $\metric$-decoding; the converse is not true. With the same method developed in the paper, it can be shown that the  type conflict error probability over the channel $W$ goes to $1$ exponentially for $R>C(W)$; even with a genie-aided  ML decoder knowing the exact conditional type $\hat{\pv}_{\yv|\xv(m)}$, the error probability would still tend to $1$ exponentially above capacity. 
	
	%{\bf AGif: define type classes $\hat P$ with distributions and not vectors/matrices $\hat \pv$}
	%%%%%%%%%%%%%%%%%%%%%%%%%%%%%
	\section{Main Result} \label{smainresult}
	In this section, we introduce the main result and discuss some of its properties.
	Our bound is derived for the maximal probability of error. Recall that for the mismatched decoding problem, a converse for the maximal probability of error implies a converse for the average probability of error \cite{csiszarDMC}.
	\begin{theorem} \label{maintheorem}
		Let $W,q$ be channel and decoding metric, respectively. We define $\bar R_\metric(W)$ as follows,
		\begin{align} 
			\bar R_\metric(W) = \max_{ P_X}\min_{\substack{P_{Y\hat{Y}|X} \in \Mc_{\rm max}(\metric)\\ P_{Y|X} = W}} I(P_X,P_{\hat{Y}|X}).
			\label{eq:upper_bound}
		\end{align}
		If $R > \bar R_\metric(W)$,  $\exists n_0 \in \NN$ and $\bar E_\metric(R)>0$ such that for $n>n_0$, the error probability of any codebook $\Cc_n$ of length $n$ and $M\geq2^{nR}$ codewords satisfies
		%\begin{align}
		$P_{e,\rm max}^\metric(\Cc_n) \geq 1 - 2 ^{-n\bar E_\metric(R)}.$
		%	\end{align}
	\end{theorem}

	\textit{{Proof Outline}:}
	The main idea behind the proof of Theorem \ref{maintheorem} is that of lower-bounding the error probability of a codebook $\Cc_n$ with $\metric$-decoding over the channel $W$ by that of the same codebook over a different channel $V$ with ML decoding, with $V=P_{\hat{Y}|X}$ as per the theorem statement. The proof is developed over the next sections of the paper. The following is an overview of the structure of the proof and the sections covering the proof.
	\begin{itemize}
		\item  In Section \ref{graph} we construct a graph $\Gc$ in the output space such that if ML decoding over $V$ makes a type conflict error for some $\yv \in \Yc^n$, then, the $\metric$-decoder makes an error for some $\hat{\yv} \in \Yc^n$ connected to $\yv$ in $\Gc$.
		\item In Section \ref{maximal} we prove a theorem that relates the maximum-likelihood decoding errors on a constructed auxiliary channel $V$ and  mismatched decoding errors on channel $W$ via the graph constructed in Section \ref{graph}. 
		\item In Sections \ref{continuous} and \ref{sec:end} we generalize the results we have derived using the method of types in the previous sections to distributions. We do this by taking the limit when $n$ tends to infinity and complete the proof of the Theorem \ref{maintheorem}.
	\end{itemize}
	\hfill$\blacksquare$

	Theorem \ref{maintheorem} implies that $C_\metric(W) \leq \bar R_\metric(W)$.  
	It is implied in Theorem \ref{maintheorem} that for any $P_{Y\hat{Y}|X} \in \Mc_{\rm max}(\metric)$ such that $P_{Y|X} = W$, 
	\begin{align}
		\bar R_\metric(W) \leq C(P_{\hat{Y}|X}).
	\end{align}
	This result is derived by using the max-min inequality:
	\begin{align}
		\bar R_\metric(W) &=\max_{P_X}\min_{\substack{P_{Y\hat{Y}|X} \in \Mc_{\rm max}(\metric)\\ P_{Y|X} = W}} I(P_X,P_{\hat{Y}|X})\\ \label{eq:maxmin_pre}
		&\leq \min_{\substack{P_{Y\hat{Y}|X} \in \Mc_{\rm max}(\metric)\\ P_{Y|X} = W}}\max_{P_X} I(P_X,P_{\hat{Y}|X}) \\
		&= \min_{\substack{P_{Y\hat{Y}|X} \in \Mc_{\rm max}(\metric)\\ P_{Y|X} = W}} C(P_{\hat{Y}|X}). 
	\end{align}
	%We thus get
	%\begin{align}
	% C_\metric(W) \leq C(P_{\hat{Y}|X}).
	%\end{align}
	As it will be shown in Section \ref{convexity}, Eq. \eqref{eq:maxmin_pre} actually holds with equality. 
	Moreover, 
	Theorem \ref{maintheorem} characterizes a family of bounds to the mismatch capacity, not only the minimum in \eqref{eq:upper_bound}. The above inequality is helpful to construct bounds without necessarily performing the optimization. 
	As an instance of the above result, setting $Y$ such that $P_{Y|X} = W$ and $\hat{Y} = Y$ makes $P_{YY|X}$ a maximal joint conditional distribution (Def. \ref{fewer4tqw}). Therefore, $C_\metric(W)\leq C(P_{Y|X})=C(W)$. In the proof it is evident that the bound remains valid for any fixed input distribution, not only the maximizing one. This means that any constant-composition codebook with type approaching a fixed $P_X$ will have an error probability that tends to one exponentially if its rate is such that
	\beq
	R>\min_{\substack{P_{Y\hat{Y}|X} \in \Mc_{\rm max}(\metric)\\ P_{Y|X} = W}} I(P_X,P_{\hat{Y}|X}).
	\eeq

	\begin{remark}
		The optimization \eqref{eq:upper_bound} in Theorem \ref{maintheorem}, is a convex-concave optimization problem. See Section \ref{convexity} for further details. 
	\end{remark}
	\begin{remark}
		It was shown \cite{csiszarDMC} that the achievability bounds for DMCs could be improved by considering an equivalent metric $\tilde\metric(x,y) = s\metric(x,y) + a(x)+b(y)$. Here we show that our bound in Theorem \ref{maintheorem} does not change by replacing metric $\metric(x,y)$ by $\tilde\metric(x,y) = s\metric(x,y) + a(x)+b(y)$. According to the definition of $\Sc_{\tilde\metric}(k_1,k_2)$, we have 
		\begin{align}
			&\argmax_{j \in \Xc} \tilde\metric(j,k_2)-\tilde\metric(j,k_1) \notag \\
			&~~~=\argmax_{j \in \Xc} 		\big(s\metric(j,k_2)+a(j)+b(k_2))\notag\\
			&~~~~~~~~~~~~~~~~~-(s\metric(j,k_1)+a(j)+b(k_1)\big) \\
			&~~~= \argmax_{j \in \Xc} s\big(\metric(j,k_2) - \metric(j,k_1)\big) + b(k_2) - b(k_1)\\
			&~~~= \argmax_{j \in \Xc} \metric(j,k_2) - \metric(j,k_1)
		\end{align}
		which is precisely the condition in the definition of $\Sc_\metric(k_1,k_2)$.
		%Therefore the sets $\Sc_\metric(k_1,k_2)$ do not change when the metric $d$ is changed to metric $\tilde{d}$.
	\end{remark}
	
	The above property from \cite{csiszarDMC}  implies that for binary-input channels, the mismatch capacity $C_\metric(W)$ is only a function of the metric differences $\metric(1,y) - \metric(2,y)$ for every $y \in \Yc$. In the remainder of this section, we show a sufficient condition for $C_\metric(W) < C(W)$ for binary-input channels based on the above observation.
	
	\begin{definition}
		We say that two sequences $\{\alpha_i\}_{i=1}^{K}$ and $\{\beta_i\}_{i=1}^{K}$ have the same order if for all $1 \leq i_1,i_2 \leq K$
		\begin{align}
			\alpha_{i_1} \geq \alpha_{i_2} \Rightarrow \beta_{i_1} \geq \beta_{i_2}.
		\end{align}
	\end{definition}

	We have the following result for $J=2$.
	\begin{theorem}\label{odjehrf7erqwe}
		Assume that  $W(k|j) > 0$, for all $j=1,2,\,k=1,\dotsc, K$. If the sequences $\big\{\log W(k|1) - \log W(k|2)\big\}_{k=1}^{K}$ and $\big\{\metric(1,k) - \metric(2,k)\big\}_{k=1}^{K}$ do not have the same order, then $C_\metric(W) < C(W)$.
	\end{theorem}
	
	\begin{IEEEproof}
		See Appendix A for the proof.
	\end{IEEEproof}
	
	\subsection{Examples}
	In the following, we discuss the applicability of our upper bound to two relevant cases. First, we show that our bound recovers known results on binary-input binary-output channels. Next, we show that our bound makes a non-trivial improvement over the channel-metric combination used in \cite{counterexample} to state the counterexample to Balakirsky's result \cite{balakirsky1995converse}.
	
	\begin{example}[Binary-input binary-output channels] \label{examp1}
		Suppose that the channel and decoding metric matrices of binary-input binary-output channels are given by
		\begin{align}
			\Wm = \begin{bmatrix}
				a  &b \\
				c  &d
			\end{bmatrix} ~~\text{and}~~ \Qm = \begin{bmatrix}
				\hat{a}  &\hat{b} \\
				\hat{c}  &\hat{d}
			\end{bmatrix}.
		\end{align}
		Without loss of generality we assume $a+d \geq b+c$. We show the following known result \cite{csiszarDMC}: if $\hat{a}+\hat{d} < \hat{b}+\hat{c}$ then $\bar R_\metric(W) = 0$. On the other hand, if $\hat{a}+\hat{d} \geq \hat{b}+\hat{c}$, then $\bar R_\metric(W) = C(W)$.
		
		\ul{\em Case 1}: $\hat{a}+\hat{d} < \hat{b}+\hat{c}$
		
		We chose the joint conditional distribution in Table \ref{tbl:maximal_example1}.
		\begin{table}[ht]
			\centering
			\caption{Joint conditional distribution $P_{Y\hat{Y}|X}$ for Example 1}
			\label{tbl:maximal_example1}
			\begin{tabular}{llll}
				\toprule
				$(k_1,k_2|j)$ & $P_{Y\hat{Y}|X}$ & $(k_1,k_2|j)$ & $P_{Y\hat{Y}|X}$ \\
				\midrule
				$(1,1|1)$ & $a - r_1$ & $(2,2|2)$ & $d - r_2$ \\
				$(1,2|1)$ & $r_1$ & $(2,1|2)$ & $r_2$ \\
				$(2,2|1)$ & $b$ & $(1,1|2)$ & $c$ \\
				$(2,1|1)$ & 0 & $(1,2|2)$ & 0 \\
				\bottomrule
			\end{tabular}
		\end{table}

		It can be checked that indeed it is a valid joint conditional distribution for $ 0 \leq r_1 \leq a$ and $ 0 \leq r_2 \leq d$, and that $\sum_{k_2}P_{Y\hat{Y}|X}(k_1,k_2|j) = P_{Y|X}(k_1|j) = W(k_1|j)$. In order to check its maximality, we first notice that for $k_1=k_2$ we always have that $\metric(i,k_2)-\metric(i,k_1) = 0$ for all $i \in \mathcal{X}$, implying that $\Sc_\metric(k_1,k_2) = \{1,2\}$. Thus, since every $j\in \mathcal{X}$ is such that $j\in \Sc_\metric(k_1,k_2) $, the corresponding four entries can be nonzero. 
		As for entry $(1,1,2)$ (resp. $(2, 2, 1)$), using the assumption $\hat{a}+\hat{d} < \hat{b}+\hat{c}$ we have that $\Sc_\metric(k_1,k_2) = \{1\}$ (resp . $\Sc_\metric(k_1,k_2) = \{2\}$), and thus they both can be nonzero. Since by assumption $\hat{a}+\hat{d} < \hat{b}+\hat{c}$, it can be checked that for entry $(2,1,2)$, $\Sc_\metric(k_1,k_2) = \{1\}$, and thus we must have $P_{Y\hat{Y}|X}(k_1,k_2|j)= 0$. Similarly for entry $(1,2,1)$, $\Sc_\metric(k_1,k_2) = \{2\}$. 
		Marginalizing the above over $Y$ gives
		\begin{align}
			P_{\hat{Y}|X} = \begin{bmatrix}
				a-r_1 \ &b+r_1 \\
				c+r_2 \ &d-r_2
			\end{bmatrix}.
		\end{align}
		Without loss of generality assume that $a$ is the largest element of $\Wm$. By setting $r_1 = r_2 = \frac{a-c}{2} = \frac{d - b}{2}$ we obtain
		\begin{align}
			P_{\hat{Y}|X} = \begin{bmatrix}
				\frac{a+c}{2}  &\frac{b+d}{2} \\
				\frac{a+c}{2}  &\frac{b+d}{2}
			\end{bmatrix}.
		\end{align} 
		Since $C(P_{\hat{Y}|X}) = 0$, we have that $C_\metric(W) \leq 0$.
		
		\ul{\em Case 2}:  $\hat{a}+ \hat{d} \geq \hat{b}+\hat{c}$
		
		In \cite{csiszargraph} it is shown that the LM achievable rate is equal to $C(W)$. Therefore, our upper-bound also matches the achievable rate.
	\end{example}

	\begin{example}\label{examp2}
		We consider the channel and metric studied in \cite{counterexample} to show a counterexample to \cite{balakirsky1995converse}
		\begin{align}
			\hspace{-2mm}\Wm = \begin{bmatrix}
				0.97 &0.03  &0\\
				0.1  &0.1   &0.8
			\end{bmatrix}
			~~\text{and}~~
			\Qm = \begin{bmatrix}
				1  &1   &1\\
				1  &0.5 &1.36
			\end{bmatrix}.
		\end{align}	 
		In this case, the LM rate is $R^{\textsc{lm}}_\metric(W) = 0.1975$ while the rate achieved by a multiletter extension of order $\ell=2$ of superposition coding gives $R^{\textsc{sc},(2)}_\metric(W)=0.1991$  \cite{counterexample} .

		We choose the maximal $P_{Y\hat{Y}|X}$ in Table \ref{tbl:maximal_example2} such that $P_{{Y}|X}=W$, which happens to be the optimal one (see Section \ref{sec:comp} for details).
		By marginalizing over $Y$ we find that 
		\begin{align}
			P_{\hat{Y}|X} = \begin{bmatrix}
				0.5 &0.5  &0\\
				0.1  &0.1   &0.8
			\end{bmatrix}.
		\end{align}	 	  
		We obtain that $\bar R_\metric(W) =0.6182$ bits/use, while the capacity is $C(W)=0.7133$ bits/use.
		\begin{table}[h]
			\centering
			\caption{Nonzero entries of $P_{Y\hat{Y}|X}$ for Example 2}
			\label{tbl:maximal_example2}
			\begin{tabular}{llll}
				\toprule
				$(k_1,k_2|j)$ & $P_{Y\hat{Y}|X}$ & $(k_1,k_2|j)$ & $P_{Y\hat{Y}|X}$ \\
				\midrule
				$(1,1|1)$ & $0.5$ & $(1,1|2)$ & $0.1$ \\
				$(1,2|1)$ & $0.47$ & $(2,2|2)$ & $0.1$ \\
				$(2,2|1)$ & $0.03$ & $(3,3|2)$ & $0.8$ \\
				\bottomrule
			\end{tabular}
		\end{table}

		In the above example, if we change $\metric(2,2)$ from $0.5$ to $1$, the same $P_{Y\hat{Y}|X}$ in Table \ref{tbl:maximal_example2} remains maximal (and optimal) and gives $\bar R_\metric(W\rev{)}=0.6182$ bits/use, matching the LM rate \cite{csiszargraph}. 
	\end{example} 
	
	\begin{example}\label{examp3}
		In this example we apply our bound to the erasures-only or zero-undetected error capacity problem. In this setting, the decoder chooses a codeword $\xv$ in the codebook if it is the only codeword with $W(\yv|\xv) >0$. Otherwise the decoder declares an erasure. The erasures-only capacity $C_{\rm eo}(W)$ is defined as the maximum achievable rate where the probability of erasure could tend to zero by increasing the block-length. It can be shown \cite{csiszarDMC} that the erasures-only capacity problem can be reduced to a mismatched decoding problem with decoding metric
		
		\begin{align} \label{defmetr}
			\metric(x,y) = \begin{cases}
				0 \ \ &W(y|x)>0 \\
				-1 \ \ &W(y|x) = 0.
			\end{cases}
		\end{align}
		In order to explain the structure of the of the sets $\Sc_\metric(k_1,k_2)$ , observe that for any two $k_1,k_2 \in \Yc$, there are two different possibilities:
		\begin{enumerate}
			\item Firstly, if there exists $j \in \Xc$ such that, $W(k_1|j) = 0$ and $W(k_2|j) > 0$ then from the definition of the metric in \eqref{defmetr} we get $j \in \Sc_\metric(k_1,k_2)$. Moreover, for any other $j' \in \Sc_\metric(k_1,k_2)$ we should have $W(k_1|j') = 0$ and $W(k_2|j') > 0$. Thus, for any $j'  \in \Sc_\metric(k_1,k_2)$ from the definition of maximality $P_{Y\hat Y|X}(k_1,k_2|j)$ could potentially be non-zero. Yet, since $P_{Y\hat Y|X}(k_1,k_2|j') \leq W(k_1|j')$, we have that $P_{Y\hat Y|X}(k_1,k_2|j')=0$. 
			\item Instead, if there is no $j \in \Xc$ such that $W(k_1|j) = 0$ and $W(k_2|j) > 0$, then
			\begin{align}
				\{j \in \Xc | W(k_2|j)>0 \} \subseteq \{j \in \Xc | W(k_1|j)>0 \}.
			\end{align}
			If $\{j \in \Xc | W(k_1|j)>0 \} = \{j \in \Xc | W(k_2|j)>0 \}$, then, outputs $k_1$ and $k_2$ can be merged without affecting $C_{\rm eo}(W)$ $[15]$. Otherwise, outputs $k_1$ and $k_2$ cannot be merged.
		\end{enumerate}
		
		Consider the following ternary-input quaternary-output channel that cannot be simplified by merging,
		\begin{align}
			\Wm = \begin{bmatrix}
				0.25 \ &0 \ &0.05 \ &0.7 \\
				0.3  \ &0.55 \ &0  \  &0.15 \\
				0.05  \  &0.5  \  &0.45 \ &0
			\end{bmatrix}.
		\end{align}
		The Shannon capacity of $W$ is $C(W) = 0.7854$ bits/use and our upper bound gives $\bar R_\metric(W) = 0.6232$ bits/use. The LM rate computed by an exhaustive search over the input distributions is $R_\metric^{\textsc{lm}}(W)=0.4292$  bits/use.	
		
		As observed from the above examples, our bound non-trivially improves on the on the trivial upper bound stating that the
mismatch capacity is at most the Shannon capacity.
		%		\begin{figure}[h]
		%			\centering
		%			\begin{tikzpicture}
		%					
		%			\draw (0,0)--node[above]{}(4,0);
		%			\draw (0,0)--node[above]{}(4,-2.6);
		%			\draw (0,-1.3)--node[above]{}(4,0);
		%			\draw (0,-1.3)--node[above]{}(4,-1.3);
		%			\draw (0,-2.6)--node[below]{}(4,0);
		%			\draw (0,-2.6)--node[below]{}(4,-1.3);
		%			\draw (0,-2.6)--node[below]{}(4,-2.6);
		%			
		%			\draw (0,0)--node[above]{}(4,-3.9);
		%			\draw (0,-1.3)--node[above]{}(4,-3.9);
		%			
		%			\filldraw [blue] (0,0) circle (2pt);
		%			\filldraw [blue] (0,-1.3) circle (2pt);
		%			\filldraw [blue] (0,-2.6) circle (2pt);
		%			\filldraw [blue] (4,0) circle (2pt);
		%			\filldraw [blue] (4,-1.3) circle (2pt);
		%			\filldraw [blue] (4,-2.6) circle (2pt);
		%			\filldraw [blue] (4,-3.9) circle (2pt);
		%			\end{tikzpicture}
		%		\end{figure}

	\end{example}

	%%%%%%%%%%%%%%%%%%%%%%%%%%%%%%%%%
	
	\section{Graph Construction}\label{graph}
	
	In this section, we outline how to construct a graph between two different conditional types obtained from a joint conditional type. 
	
	\begin{definition}\label{def:graph1}
		Let $\Gc = \{\Vc_1,\Vc_2,\Ec\}$ be a regular bipartite graph with vertex sets $\Vc_1$ and $\Vc_2$, edge set $\Ec$ and 
		degrees $r_1$ on vertex set $\Vc_1$ and $r_2$ on vertex set $\Vc_2$. 
		For $\Bc \subset \Vc_2$ we define the set of vertices in $\Vc_1$ connected to $\Bc$ as
		\begin{align}\label{sidef}
			\Psi_{21}(\Bc) = \big\{v \in \Vc_1\,|\,\exists b \in \Bc;  (b,v) \in \Ec\big\} .
		\end{align}
	\end{definition} 
	Analogously for $\Bc \subset \Vc_1$, the set $\Psi_{12}(\Bc)$ is defined similar to \eqref{sidef}.
	\begin{lemma}\label{lemmafhghgh}
		Suppose $\Gc = \{\Vc_1,\Vc_2,E\}$ is a regular bipartite graph with degrees $r_1>0,r_2>0$. Then, for any $\Bc \subset \Vc_2$ we have that
		\begin{align} \label{grapheq}
			\frac{|\Psi_{21}(\Bc)|}{|\Vc_1|} \geq \frac{|\Bc|}{|\Vc_2|}.
		\end{align}
	\end{lemma}
	
	\begin{IEEEproof}
		Let $\Bc \subset \Vc_2$ and consider $\Psi_{21}(\Bc)$. There are exactly $r_2|\Bc|$ edges between $\Bc$ and $\Psi_{21}(\Bc)$. Since each vertex in $\Psi_{21}(\Bc)$ is connected to at most $r_1$ vertices of $\Bc$ we have
		\begin{align}
			r_2|\Bc| \leq r_1|\Psi_{21}(\Bc)|
			\label{eq:r2r1}
		\end{align}	
		which implies that
		\begin{align}
			\frac{r_2}{r_1}|\Bc| \leq |\Psi_{21}(\Bc)|.
			\label{eq:r2r12}
		\end{align}
		Since there are exactly $r_1|\Vc_1| = r_2|\Vc_2|$ edges in the graph, the result follows by substituting $\frac{r_2}{r_1} = \frac{|\Vc_1|}{|\Vc_2|}$ in \eqref{eq:r2r12}.
	\end{IEEEproof}
	
	%%%%%%%%%%%%%%%%%%%%%%%%%%%%%%%%%%%%%%%%%%%%
	%\section{Constructing a graph}\label{mapping}
	Our aim is to construct a graph between different two conditional type classes, in order to be able to relate type conflict errors of codebook $\Cc_n$ over the channel $V$ and errors of $\Cc_n$ over the channel $W$ under $\metric$-decoding. Suppose $\pv_{Y\hat{Y}|X} \in \Pc_n(\Yc\times\hat{\Yc}|\Xc)$ is an arbitrary joint conditional type. We construct a graph between $\Tc_{\xv}^n(\pv_{Y|X})$ and $\Tc_{\xv}^n(\pv_{\hat{Y}|X})$, the corresponding conditional type classes.
	
	\begin{definition}\label{graphconstruction}
		The graph 
		\beq
		\Gc_\xv(\pv_{Y\hat{Y}|X}) = \big\{\Tc_{\xv}^n(\pv_{Y|X}),\Tc_{\xv}^n(\pv_{\hat{Y}|X}), \Ec\big\}
		\eeq has the following edge set:
		\begin{align}\label{5tgpoetyo}
			\Ec = \big\{(\yv,\hat{\yv}) \,|\,\hat{\pv}_{\yv\hat{\yv}|\xv} = \pv_{Y\hat{Y}|X}\big\}.
		\end{align}	
	\end{definition}
	
	\begin{lemma}\label{graphlemma}
		The graph $\Gc_\xv(\pv_{Y\hat{Y}|X})$ is  regular, i.e. all sequences in each conditional type class $\Tc_\xv^n(\pv_{Y|X})$ and
		$T^n_\xv(\pv_{\hat{Y}|X})$ have the same degree.
	\end{lemma}
	\begin{IEEEproof}
		For a given $\xv\in \Tc^n(\pv_X)$, $ |\Tc_{\xv}^n(\pv_{Y|X})|$ is independent of the chosen $\xv\in \Tc^n(\pv_X)$, but dependent on $\pv_X$. Similarly, for a given $\yv \in \Tc_{\xv}^n(\pv_{Y|X})$, $|\Tc^n_{\yv\xv}(\pv_{Y\hat{Y}|X})|$ is independent of the chosen $\xv,\yv$, but dependent on the joint type $\pv_{XY}$. Therefore, the total number of edges that are connected to any given $\yv \in \Tc_{\xv}^n(\pv_{Y|X})$ is  equal to $|\Tc^n_{\yv\xv}(\pv_{Y\hat{Y}|X})|$ (see \eqref{lfnrnfgrn}). This proves the left-regularity, \textit{i.e.}, for vertex set $\Tc_{\xv}^n(\pv_{Y|X})$. The same argument holds for $\hat{\yv} \in \Tc_{\xv}^n(\pv_{\hat{Y}|X})$ and therefore the graph is regular.
	\end{IEEEproof}
	
	As we show next, the combination of Lemmas \ref{lemmafhghgh} and \ref{graphlemma} will prove to be helpful. Assume for a codeword $\xv$ we find a set $\Bc \subset \Tc_{\xv}^n(\pv_{\hat Y|X})$ that yields a type conflict error (see Definition \ref{bodsjgawdfabwd}). Then, the probability of an element $\hat \yv \in \Bc$ being  the output of an arbitrary channel $V$ given that the conditional type is $\pv_{\hat Y|X}$, is given by
	\begin{align}\label{grrrwer}
		\PP\big[\hat \yv \in \Bc\,|\,\hat\yv \in \Tc_{\xv}^n(\pv_{\hat Y|X}),\xv \text{\ is sent}\big] = \frac{|\Bc|}{|\Tc_{\xv}^n(\pv_{\hat{Y}|X})|}
	\end{align}  
	where the probability is computed with respect to an auxiliary memoryless channel $V$, i.e., $\PP \big[\hat \yv | \xv \text{\ is sent}\big] = \prod_{i=1} V(\hat y_i | x_i)$ and equality holds because all elements of $\Tc_{\xv}^n(\pv_{\hat{Y}|X})$ are equally likely to appear at the output when $\xv$ is sent. The probability in \eqref{grrrwer} should be understood as the probability of the set $\Bc$ given that $\hat\yv \in \Tc_{\xv}^n(\pv_{\hat Y|X})$ and $\xv \text{\ is sent}$. Therefore, if the graph $\Gc_\xv(\pv_{Y\hat{Y}|X})$ is connecting $\hat \yv$ causing a type conflict error to $\yv$ causing a $\metric$-decoder error, by Lemma \ref{lemmafhghgh} we show that the set $\Psi_{21}(\Bc)\subset \Tc_{\xv}^n(\pv_{{Y}|X})$ satisfies
	\begin{align}\label{oerfgrh}
		\frac{|\Psi_{21}(\Bc)|}{|\Tc_{\xv}^n(\pv_{Y|X})|} \geq \frac{|\Bc|}{|\Tc_{\xv}^n(\pv_{\hat{Y}|X})|}.
	\end{align}
	Using the same argument as in \eqref{grrrwer} we have
	\begin{align}
		\PP\big[\yv \in \Psi_{21}(\Bc)\,|\,\yv \in \Tc_{\xv}^n(\pv_{Y|X}),\xv \text{\ is sent}\big] = \frac{|\Psi_{21}(\Bc)|}{|\Tc_{\xv}^n(\pv_{Y|X})|}.
		\label{eq:psipsi}
	\end{align}
	Combining \eqref{eq:psipsi} and \eqref{oerfgrh} we get
	\begin{align} 
		\PP[\yv \in \Psi_{21}(\Bc)\,|\,\yv \in &\Tc_{\xv}^n(\pv_{Y|X}),\xv \text{\ is sent}] \notag \geq \\
		& \PP[\hat \yv \in \Bc\,|\,\hat\yv \in \Tc_{\xv}^n(\pv_{\hat Y|X}),\xv \text{\ is sent}].
	\end{align}
	As a result, we get a lower bound on the probability of error of the $\metric$-decoder in channel $W$ as a function of type conflict errors in channel $V$. In the next section, we prove that a graph constructed based on a  maximal joint conditional type has the property of connecting type conflict  errors to $\metric$-decoder errors.
	
	%%%%%%%%%%%%%%%%%%%%%%%%%%%%%%%%%%%
	\section{Connecting $\metric$-decoding Errors and Type Conflict Errors} \label{maximal} 
	
	We next introduce a property of maximal joint conditional types and use it to relate type conflict and $\metric$-decoding errors.
	\begin{lemma}\label{lemmaximal}
		Let $\pv_X \in \Pc_n(\Xc)$, $\xv,\hat{\xv} \in \Tc^n(\pv_X)$, and $\pv_{Y\hat{Y}|X}$ be a maximal  joint conditional type. If $\hat{\yv} \in \Tc_{\xv}^n(\pv_{\hat{Y}|X}) \cap \Tc_{\hat{\xv}}^n(\pv_{\hat{Y}|X})$ is connected to $\yv \in \Tc_{\xv}^n(\pv_{Y|X})$ in $\Gc_\xv(\pv_{Y\hat{Y}|X})$ then,
		\begin{align}
			\metric(\xv,\yv) \leq \metric(\hat{\xv},\yv).\label{geewt4}
		\end{align}
		
	\end{lemma}
	
	\begin{IEEEproof}
		From the definition of type, for any $\bar{\xv} \in \Xc^n$,
		\begin{align}
			\hat{\pv}_{\yv\hat{\yv}}(k_1,k_2) = \sum_{j}\hat{\pv}_{\bar{\xv}\yv\hat{\yv}}(j,k_1,k_2).  \label{equivalent}
		\end{align}
		We use the above equation once by setting $\bar{\xv} = \xv$ and once by setting $\bar{\xv} = \hat{\xv}$. Therefore, we have
		\begin{align}
			\sum_{j}\hat{\pv}_{\xv\yv\hat{\yv}}(j,k_1,k_2) = \sum_{j}\hat{\pv}_{\hat{\xv}\yv\hat{\yv}}(j,k_1,k_2).\label{rtat4}
		\end{align}
		We continue by bounding $\metric(\hat{\xv},\hat{\yv}) - \metric(\hat{\xv},\yv)$ as
		\begin{align}
			&\metric(\hat{\xv},\hat{\yv}) - \metric(\hat{\xv},\yv) \notag\\
			&= n\sum_{j,k_1,k_2}\hat{\pv}_{\hat{\xv}\yv\hat{\yv}}(j,k_1,k_2)\big(\metric(j,k_2) - \metric(j,k_1)\big)\label{werqt}\\ \label{ld;ejjreir}
			&\leq n\sum_{k_1,k_2}\Big(\sum_{j}\hat{\pv}_{\hat{\xv}\yv\hat{\yv}}(j,k_1,k_2)\Big)\max_{j'}\big(\metric(j',k_2) - \metric(j',k_1)\big) \\ \label{sa;efq}
			&= n\sum_{k_1,k_2}\Big(\sum_{j}\hat{\pv}_{\xv\yv\hat{\yv}}(j,k_1,k_2)\Big)\max_{j'}\big(\metric(j',k_2) - \metric(j',k_1)\big)  \\ \label{fwefne}
			&= n\sum_{k_1,k_2}\sum_{j}\hat{\pv}_{\xv\yv\hat{\yv}}(j,k_1,k_2)\big(\metric(j,k_2) - \metric(j,k_1)\big) \\ \label{fr4werq}
			&= \metric(\xv,\hat{\yv}) - \metric(\xv,\yv)
		\end{align}
		where \eqref{werqt} follows  from the definition of metric and type, since for a joint type $\hat{\pv}_{\xv\yv}$ we have that $\metric(\xv,\yv)=n\sum_{j,k}\hat{\pv}_{\xv\yv}(j,k)\metric(j,k)$, \eqref{ld;ejjreir} follows from upper-bounding $(\metric(j,k_2) - \metric(j,k_1))$ by $\max_{j}(\metric(j,k_2) - \metric(j,k_1))$, \eqref{sa;efq} follows from \eqref{rtat4}, \eqref{fwefne} follows from the maximality of $\pv_{Y\hat{Y}|X}$ (see Definition \eqref{fewer4tqw}) and the graph construction $\Gc_\xv(\pv_{Y\hat{Y}|X})$ (see Definition \eqref{graphconstruction}) and \eqref{fr4werq} follows again from the metric definition. 
		
		Using the fact that $\hat{\yv} \in \Tc_{\xv}^n(\pv_{\hat{Y}|X}) \cap \Tc_{\hat{\xv}}^n(\pv_{\hat{Y}|X})$ and since the types of $\xv$ and $\hat \xv$ are the same, we get a type conflict error, \textit{i.e.}, $\hat{\pv}_{\hat{\yv}|\xv} = \hat{\pv}_{\hat{\yv}|\hat{\xv}}$. Thus, $\metric(\xv,\hat{\yv}) = \metric(\hat{\xv},\hat{\yv})$. Finally, combining with \eqref{fr4werq} we get the desired result $\metric(\xv,\yv) \leq \metric(\hat{\xv},\yv)$, \textit{i.e.}, a $\metric$-decoding error. See Appendix C for the case where the decoding metric  $\metric$ takes $-\infty$ values.
	\end{IEEEproof}

	The above lemma states that if $\hat \yv \in \Tc_{\xv}^n(\pv_{\hat{Y}|X}) \cap \Tc_{\hat{\xv}}^n(\pv_{\hat{Y}|X})$ and if $\xv,\hat{\xv}\in \Cc_n$, by observing $\hat \yv$ when $\xv$ is sent, there will be a type conflict error. Moreover, if such a $\hat \yv$ is connected to $\yv$ in $\Gc_\xv(\pv_{Y\hat{Y}|X})$, then, based on \eqref{geewt4}, by observing $\yv$ when $\xv$ is sent, the $\metric$-decoder makes an error. 
	
	\begin{definition} \label{sumset}
		Let $W$ be a channel and $\pv_X\in \Pc_n(\Xc)$ an input type. We define the channel type neighborhood as the set of conditional types that are close to $W$,

		\begin{align} \label{mdakaodapwpr}
			\Nc_{\epsilon,\pv_X}(W) = \big\{&\pv_{Y|X} \in \Pc_n({\Yc|\Xc}) \,|\,\, \forall j,k ~\text{if}~ \pv_{X}(j)>0, \notag\\
			&~~~~~|W(k|j) - \pv_{Y|X}(k|j)| \leq \epsilon\big\}.
		\end{align}
		
	\end{definition}

	The previous result showed that if $\pv_{Y\hat{Y}|X}$ is a maximal joint conditional type, then type conflict errors in $\Tc_{\xv}^n(\pv_{\hat{Y}|X})$ can be related to $\metric$-decoding errors in $\Tc_{\xv}^n(\pv_{Y|X})$.
	Assume $P_{Y\hat{Y}|X}$ is a maximal joint conditional distribution such that $P_{Y|X} = W$ and $P_{\hat{Y}|X} = V$. The lemma below shows that for every empirical conditional type $\overline{\Wm}$ close to the channel $W$ there exists a maximal joint conditional type, that can be used to relate type conflict errors of a type close to $V$ to $\metric$-decoder errors over $W$.

	\begin{lemma}\label{rahat}
		Let $\pv_X \in \Pc_n(\Xc)$ be an input type and $\pv_{\rm min} \eqdef \min_{j,\pv_X(j) > 0} \pv_X(j)$. Assume $P_{Y\hat{Y}|X}$ is a maximal joint conditional distribution such that $P_{Y|X} = W$ and $P_{\hat{Y}|X} = V$. Moreover, let $\epsilon \geq \frac{2K}{n\pv_{\rm min}}$. Then, for each $\overline{\Wm}\in\Nc_{\frac{\epsilon}{2},\pv_X}(W)$, we can find a maximal joint conditional type $\bar\pv_{Y\hat{Y}|X}$ such that $\bar\pv_{Y|X} = \overline{\Wm}$ and $\bar\pv_{\hat{Y}|X} \in \Nc_{2K\epsilon,\pv_X}(V) $.
	\end{lemma}
	\begin{IEEEproof}
		If $\epsilon > 1$ then there is nothing to prove. Therefore, we consider $\epsilon < 1$, For $j\in\Xc$ and $k_1,k_2\in\Yc$, choose $\pv_{Y\hat{Y}|X}(k_1,k_2|j)$ to be either
		
		\begin{align}\label{tegojrgrj}
			\pv_{Y\hat{Y}|X}(k_1,k_2|j) = \frac{\big \lfloor n\pv_X(j)P_{Y\hat{Y}|X}(k_1,k_2|j) \big \rfloor}{n\pv_X(j)}
		\end{align}
		or
		\begin{align}\label{f4rtopwer}
			\pv_{Y\hat{Y}|X}(k_1,k_2|j) = \frac{\big \lceil n\pv_X(j)P_{Y\hat{Y}|X}(k_1,k_2|j) \big \rceil}{n\pv_X(j)}
		\end{align}
		such that for every $j \in \Xc$ we have
		\begin{align}
			\sum_{k_1,k_2} \pv_{Y\hat{Y}|X}(k_1,k_2|j) = 1.
		\end{align}
		Such a choice is possible since
		\begin{align}
			\sum_{k_1,k_2} P_{Y\hat{Y}|X}(k_1,k_2|j) = 1.
		\end{align}
		
		Moreover, when $\pv_X(j) = 0$ define $\pv_{Y\hat{Y}|X}(k_1,k_2|j)$ as in \eqref{definetype}. The above choice implies that for every $j\in\Xc$ such that $\pv_X(j) > 0$ and any $k_1,k_2\in\Yc$,
		
		\begin{align}
			\big|\pv_{Y\hat{Y}|X}(k_1,k_2|j) - P_{Y\hat{Y}|X}(k_1,k_2|j)\big| &\leq \frac{1}{n\pv_X(j)}\\
			&\leq \frac{1}{n\pv_{\rm min}}.\label{d3ewfwegwf}
		\end{align}
		
		Moreover, based on \eqref{tegojrgrj} and \eqref{f4rtopwer} $\pv_{Y\hat{Y}|X}$ is maximal, since $\pv_{Y\hat{Y}|X}(k_1,k_2|j)$ is non-zero either when $k_1 = k_2$ or for the same entries that $P_{Y\hat{Y}|X}$ is non-zero. As a result of \eqref{d3ewfwegwf} for every $j\in\Xc$ such that $\pv_X(j) > 0$ we have that
		
		\begin{align}
			\Big| \sum_{k_2}\big( &\pv_{Y\hat{Y}|X}(k_1,k_2|j) - P_{Y\hat{Y}|X}(k_1,k_2|j)\big)\Big| \notag\\
			&\leq \sum_{k_2} \big| \pv_{Y\hat{Y}|X}(k_1,k_2|j) -  P_{Y\hat{Y}|X}(k_1,k_2|j)\big|\\
			&\leq \frac{K}{n\pv_{\rm min}}
		\end{align}
		
		and thus,
		\begin{align} \label{cpmieawnwuewa}
			\pv_{Y|X} &\in  \Nc_{\frac{K}{n\pv_{\rm min}},\pv_X}(W) \\ \label{pgtrjgtirehgtfru}
			\pv_{\hat{Y}|X} &\in \Nc_{\frac{K}{n\pv_{\rm min}},\pv_X}(V).
		\end{align}
		
		For any $\overline{\Wm} \in \Nc_{\frac{\epsilon}{2},\pv_X}(W)$, for every $j\in\Xc$ such that $\pv_X(j) > 0$ and $k \in \Yc$ by definition we know that $|\overline{\Wm}(k|j) - \pv_{Y|X}(k|j) |\leq \epsilon$, since
		
		\begin{align}
			\big|\overline{\Wm}(k|j) - \pv_{Y|X}(k|j)\big | &\leq \big|\overline{\Wm}(k|j) - W(k|j) \big| \notag \\
			&~~~~+\big| W(k|j)- \pv_{Y|X}(k|j) \big| \\ \label{wdasfdfgwf}
			&\leq \frac{\epsilon}{2} + \frac{\epsilon}{2} \\
			&= \epsilon.
		\end{align}
		
		where \eqref{wdasfdfgwf} follows from \eqref{cpmieawnwuewa} and \eqref{pgtrjgtirehgtfru}.
		Construct $\bar\pv_{Y\hat{Y}|X}$ from $\pv_{Y\hat{Y}|X}$ in the following way. For any $j,k_1 \in \Xc \times \Yc$ such that $\pv_X(j) > 0$, if $\overline{\Wm}(k_1|j) - \pv_{Y|X}(k_1|j) > 0$ add non-negative real numbers less than or equal $\epsilon$ to $\pv_{Y\hat{Y}|X}(k_1,k_2|j), \ k_2 = 1,2,\dotsc, K$ to obtain $\bar\pv_{Y\hat{Y}|X}$ with the following property,
		\begin{align} \label{ogrmfada}
			\sum_{k_2} \bar\pv_{Y\hat{Y}|X}(k_1,k_2|j) = \overline{\Wm}(k_1|j).
		\end{align}  
		We can do this because $|\overline{\Wm}(k_1|j) - \pv_{Y|X}(k_1|j) | \leq \epsilon$. Note that by construction of this step all entries of $\bar\pv_{Y\hat{Y}|X}$ so far are non-negative.
		
		We can do the same if $\overline{\Wm}(k_1|j) - \pv_{Y|X}(k_1|j) \leq 0$ with non-positive real numbers not less than $-\epsilon$ such that 
		\begin{align} \label{naifmoefa}
			\sum_{k_2} \bar\pv_{Y\hat{Y}|X}(k_1,k_2|j) = \overline{\Wm}(k_1|j).
		\end{align}  
		Observe that from $\overline{\Wm}(k_1|j) - \pv_{Y|X}(k_1|j) \leq 0$ and $|\overline{\Wm}(k_1|j) - \pv_{Y|X}(k_1|j) | \leq \epsilon$ we obtain that  $ -\epsilon \leq \overline{\Wm}(k_1|j) - \pv_{Y|X}(k_1|j) \leq 0$. For the above step,  we can perform the addition of non-negative numbers in a way that makes all the entries of $\bar\pv_{Y\hat{Y}|X}$ non-negative. This is true since we know $\sum_{k_2} \pv_{Y\hat{Y}|X}(k_1,k_2|j) = \pv_{Y|X}(k_1|j) $ and $ -\epsilon \leq \overline{\Wm}(k_1|j) - \pv_{Y|X}(k_1|j) \leq 0$.
		Then marginalizing over $\hat{Y}$ we get $\bar\pv_{\hat{Y}|X}$ satisfying the following
		\begin{align}
			&|\bar\pv_{\hat{Y}|X}(k_2|j) - \pv_{\hat{Y}|X}(k_2|j)| \notag\\
			 &= \Big|\sum_{k_1}\bar\pv_{Y\hat{Y}|X}(k_1,k_2|j) - \pv_{Y\hat{Y}|X}(k_1,k_2|j)\Big| \\
			&\leq \sum_{k_1}|\bar\pv_{Y\hat{Y}|X}(k_1,k_2|j) - \pv_{Y\hat{Y}|X}(k_1,k_2|j)|\\
			&\leq \sum_{k_1}\epsilon \\
			&= K\epsilon.
		\end{align}
		
		Therefore by the triangle inequality and \eqref{pgtrjgtirehgtfru} we get $|\bar\pv_{\hat{Y}|X}(k_2|j) - \Vm(k_2|j)| \leq 2K\epsilon$. 
	\end{IEEEproof}
	
	In the next theorem, we show that if $P_{Y\hat{Y}|X}$ is a maximal joint conditional distribution and $M$ is large enough, then we will find many type conflict errors over conditional types close to $V = P_{\hat{Y}|X}$. These are then linked to $\metric$-decoding errors over the channel $W = P_{Y|X}$. 
	
	\begin{theorem} \label{theo2}
		Let $\Cc_n$ be a codebook with $M$ codewords and composition $\pv_X$ with $\pv_{\rm min} \eqdef \min_{j,\pv_X(j) > 0} \pv_X(j)$. Let $P_{Y\hat{Y}|X}$ be a maximal joint conditional distribution such that $P_{Y|X} = W, P_{\hat{Y}|X} = V$. Let $ \epsilon \geq \frac{2K}{n\pv_{\rm min}}$ and suppose $\Nc_{2K\epsilon,\pv_X}(V) = \{\Vm^1,\Vm^2\dotsc, \Vm^t\} $. Let $\qv^i$ be the output type corresponding to input type $\pv_X$ and conditional type $\Vm^i$. If for some integer $a \geq 2$, for every $\xv \in \Tc^n(\pv_X)$ and for all $i\in\{1,\dotsc,t\}$ we have that
		\begin{align}\label{condition}
			M|\Tc_{\xv}^n(\Vm^i)| \geq a^2 (n+1)^{2J(K-1)} \max_{1\leq i' \leq t}|\Tc^n(\qv^{i'})|,
		\end{align}
		then, there exists a codeword $\xv(m) \in \Cc_n$ such that
		\begin{align}\label{refewrfeg}
			\PP\Big[\hat{m} \neq m\,\big|\, \hat{\pv}_{\yv|\xv(m)} \in\Nc_{\frac{\epsilon}{ 2},\pv_X}(W),&\xv(m) \text{\ is sent}\Big]\notag\\& > 1 - \frac{2}{a+1}.
		\end{align}
	\end{theorem}

	The above theorem gives us a sphere-packing type of bound. From the method of types we know that $|\Tc_{\xv}^n(\Vm^i)| \doteq 2^{nH(\Vm^i|\pv_X)} \approx 2^{nH(V|\pv_X)}$ and that $|\Tc^n(\qv^i)| \doteq 2^{nH(\qv^i)} \approx 2^{nH(\qv)}$, where $\qv$ denotes the output distribution induced by input type $\pv_X$ and channel $\Vm$. The approximation comes from the definition and properties of the neighborhood introduced in Definition \ref{sumset} (see Section \ref{sec:end} for more details). Therefore, inequality \eqref{condition} roughly implies that
	\begin{align}
		2^{nR}2^{nH(V|\pv_X)} \gtrsim 2^{nH(\qv)},
	\end{align} 
	or equivalently, $R \gtrsim I(\pv_X,V)$.  The theorem states that if $R \gtrsim I(\pv_X,V)$, the error probability of one of the messages is high under $\metric$-decoding.
	
	\begin{IEEEproof}
		The proof is divided into the following four parts:
		\begin{enumerate}
			\item The existence of a codeword $\xv(m)\in\Cc_n$ that yields a type conflict error with many other codewords for many output sequences
			\item An error probability lower bound for the above codeword $\xv(m)\in\Cc_n$ based on Lemma 3
			\item An overall error probability lower bound when the channel type is $\overline \Wm\in \Nc_{\frac{\epsilon}{ 2},\pv_X}(W)$
			\item An overall error probability lower bound for all channel types  in the neighborhood $\Nc_{\frac{\epsilon}{ 2},\pv_X}(W)$
		\end{enumerate}
		
		\ul{Part 1}
		
		The first step is to show that there is a codeword $\xv(m)\in\Cc_n$ such that for all $1 \leq i \leq t$ a large proportion of sequences in $\Tc_{\xv(m)}^n(\Vm^i)$ have the same conditional type $\Vm^i$ with at least $a$ other codewords in $\Cc_n$, yielding a type conflict error with these $a$ codewords. More precisely, we wish to show that there is a codeword $\xv(m)\in\Cc_n$ and a family of sets $\mathcal{F} = \big\{\Bc_i \,|\,\Bc_i \subset \Tc_{\xv(m)}^n(\Vm^i)$, $i = 1,2,\dotsc, t\big\}$ such that
		\begin{enumerate}
			\item $|\Bc_i| \geq \frac{a-1}{a}|\Tc_{\xv(m)}^n(\Vm^i)| $
			\item $\forall \hat\yv \in \Bc_i$  there are $a$ other codewords  $\xv'(1),\xv'(2),...,\xv'(a) \in \Cc_n$ for which $\hat{\pv}_{\hat\yv|\xv'(1)}=\dotsc=\hat{\pv}_{\hat\yv|\xv'(a)} = \hat{\pv}_{\hat\yv|\xv(m)} = \Vm^i $. 
		\end{enumerate}

		%	\begin{align}
		%	&1)|\Bc_i| \geq \frac{a-1}{a}|\Tc_{\xv(m)}^n(\Vm^i)| \label{Bisbig}  \\ 
		%	&2)\forall \hat\yv \in \Bc_i  there are $a$ other codewords  \xv'(1),\xv'(2),...,\xv'(a) \in \Cc \text{ which } \hat{\pv}_{\yv|\xv'(s)} = \hat{\pv}_{\yv|\xv(m)} = \Vm^i \label{Berror}\\ 
		%&\text{ for } s = 1,2,...,a. 
		%	\end{align}
		
		%\AGF{Do the type-conflict error codewords $\xv'(1),\xv'(2),...,\xv'(a) \in \Cc_n$ depend on index $i$?}
		
		This implies that we can find a family of sets $\Fc = \{\Bc_i\}$ where $\Bc_i \subset \Tc_{\xv(m)}^n(\Vm^i)$, such that all members of $\Bc_i$ for $1\leq i \leq t$ cause type conflict errors with other $a$ codewords.
		
		We prove this result by contradiction. Suppose there is no such $\xv(m)$ with such family $\Fc = \{\Bc_i\}$. Then, there is no $\xv \in \Cc_n$, such that a family $\Fc = \{\Bc_i\}$ with the above properties exists. Therefore, for any $\xv \in \Cc_n$ there is a set $\Ac_{\xv}$ with the following properties:
		\begin{enumerate}
			\item $\Ac_{\xv} \subset \Tc_{\xv}^n(\Vm^i)$ for some $1 \leq i \leq t$,
			\item $|\Ac_{\xv}| > \frac{1}{a}|\Tc_{\xv}^n(\Vm^i)|$ for the same $i$ in condition $1$ above, \label{AxT}
			\item $\forall \hat\yv \in \Ac_{\xv}$ there are at most $a-1$ other codewords $\xv'(1),\xv'(2),...,\xv'(a-1)\in \Cc_n$ such that $\hat{\pv}_{\hat\yv|\xv'(1)} = \dotsc = \hat{\pv}_{\hat\yv|\xv'(a-1)} = \hat{\pv}_{\hat\yv|\xv}$. \label{unicode}
		\end{enumerate}
		
		There are at most $(n+1)^{J(K-1)}$ conditional types $\pv_{Y|X}$ such that $\pv_{Y} = \qv^i$, and thus, $t \leq (n+1)^{J(K-1)}$. We claim that every $\hat \yv \in \Tc^n(\qv^{i})$, for any $1\leq i\leq t$ is a member of at most $a(n+1)^{J(K-1)}$ sets $\Ac_{\xv}$. In order to show this, assume that some $\hat \yv$ violates this claim and is a member of more than $a(n+1)^{J(K-1)}$ sets $\Ac_{\xv}$. Then, by the pigeonhole principle, there are at least $\left \lceil \frac{a(n+1)^{J(K-1)}+1}{(n+1)^{J(K-1)}} \right \rceil = a+1$ sets $\Ac_{\bar\xv(1)}\subset\Tc_{\bar\xv(1)}^n(\Vm^{i_1}),\dotsc,\Ac_{\bar\xv(a+1)}\subset\Tc_{\bar\xv(a+1)}^n(\Vm^{i_1})$ corresponding to codewords $\bar\xv(1),\dotsc,\bar\xv(a+1)\in\Cc_n$ for the same $ 1\leq i_1\leq t$. In the above argument pigeons are the sets $\Ac_{\xv}$ that contain $\hat \yv$ and pigeonholes are the indices $1\leq i\leq t$ of $\Tc_{\xv}^n(\Vm^{i})$ such that that $\Ac_{\xv} \subset \Tc_{\xv}^n(\Vm^{i})$. Therefore, since, $\hat\yv \in \Ac_{\bar\xv(1)} \cap \Ac_{\bar \xv(2)} \cdots \cap \Ac_{\bar\xv(a+1) }$, we have that
		\begin{align}
			\hat{\pv}_{\hat\yv|\bar\xv(1)} = \hat{\pv}_{\hat\yv|\bar\xv(2)} = \cdots = \hat{\pv}_{\hat\yv|\bar\xv(a+1)} = \Vm^{i_1}
		\end{align}
		which contradicts the third condition that the sets $\Ac_{\xv}$ must satisfy. Therefore, the claim that every $\hat \yv \in \Tc^n(\qv^{i})$, for any $1\leq i\leq t$ is a member of at most $a(n+1)^{J(K-1)}$ sets $\Ac_{\xv}$ is verified.
		%As a result, for all $1 \leq i \leq t$ each element of $\Tc^n(\qv^i)$ is an element of at most $a(n+1)^{J(K-1)}$ sets $\Ac_{\xv}$. 
		Furthermore, considering the fact that $\Ac_{\xv} \subset \Tc_{\xv}^n(\Vm^i)$ and each element of $\Tc^n(\qv^i)$ is in at most $a(n+1)^{J(K-1)}$ sets $\Ac_{\xv}$ we get the following,
		\begin{align} \label{bvefoxsmanva}
			\sum_{\xv \in \Cc_n}^{}|\Ac_{\xv}| & =  \sum_{\xv \in \Cc_n}\sum_{\hat \yv} \indicator \{\hat \yv \in \Ac_{\xv}  \} \\ 
			& = \sum_{\hat \yv}  \sum_{\xv \in \Cc_n}\indicator \{\hat \yv \in \Ac_{\xv}  \}
			\\ \label{dawfawofjiwaf}
			& \leq \sum_{\hat \yv} a(n+1)^{J(K-1)} \\ \label{cbsbawfwf}
			&= a(n+1)^{J(K-1)}\sum_{i' = 1}^{t} |\Tc^n(\qv^{i'})| \\ \label{fhawfovse}
			&	\leq a(n+1)^{J(K-1)}\cdot t \cdot \max_{1\leq i' \leq t}|\Tc^n(\qv^{i'})|
			\\ \label{gvyafkoawd}
			&= a(n+1)^{2J(K-1)}\max_{1\leq i' \leq t}|\Tc^n(\qv^{i'})|.
		\end{align}
		where \eqref{dawfawofjiwaf} follows since we have shown that each element of any $\Tc^n(\qv^{i'})$ is a member of at most $a(n+1)^{J(K-1)}$ sets $\Ac_{\xv}$,  \eqref{cbsbawfwf} follows from converting the sum over $\hat \yv$ into sum over types $\Tc^n(\qv^{i'})$. We can do this because every $\hat \yv$ is a member of $\Tc^n(\qv^{i'})$ for some $1\leq i' \leq t$ and sets $\Tc^n(\qv^{i'})$ are disjoint. Moreover, \eqref{fhawfovse} follows by upper bounding $|\Tc^n(\qv^{i'})|$ by  $\max_{1\leq i' \leq t}|\Tc^n(\qv^{i'})|$ and \eqref{gvyafkoawd} follows from $t \leq (n+1)^{J(K-1)}$.
		On the other hand,
		\begin{align}\label{Aisbig}
			\sum_{\xv \in \Cc_n}^{}|\Ac_{\xv}|  &> M\frac{1}{a}|\Tc_{\xv}^n(\Vm^i)|\\ \label{AisbigerQ}
			&\geq a(n+1)^{2J(K-1)}\max_{1 \leq i \leq t}|\Tc^n(\qv^i)|.
		\end{align}
		where \eqref{Aisbig} follows from the second property of the sets $\Ac_{\xv}$ and \eqref{AisbigerQ} from the second condition of Theorem \ref{theo2}.
		Inequalities \eqref{gvyafkoawd} and \eqref{AisbigerQ} lead to a contradiction because expressions are the same but one is strictly smaller than  $\sum_{\xv \in \Cc_n}^{}|\Ac_{\xv}|$ and the other one is larger than or equal to $\sum_{\xv \in \Cc_n}^{}|\Ac_{\xv}|$. 
		
		Therefore, we can find a codeword $\xv(m)\in\Cc_n$ with a family of sets $\Fc = \{\Bc_i\}$ such that $\Bc_i \subset\Tc_{\xv(m)}^n(\Vm^i)$ are large enough and yield type conflict errors with at least $a$ codewords.

		\ul{Part 2}
		
		We proceed by  using the assumption that $P_{Y\hat{Y}|X}$ is a maximal joint conditional distribution. Since $P_{Y\hat{Y}|X}$ is maximal, based on the Lemma \ref{rahat} for any $\overline{\Wm} \in \Nc_{\frac{\epsilon}{2},\pv_X}(W)$ we can find a maximal joint conditional type $\pv_{Y\hat{Y}|X}$ such that $\pv_{Y|X} = \overline{\Wm}$ and $\pv_{\hat{Y}|X} = \Vm^i \in \Nc_{\epsilon,\pv_X}(V)$. Construct the graph $\Gc_{\xv(m)} (\pv_{Y\hat{Y}|X})$ for the codeword $x(m)$ we found above, connecting $\Tc_{\xv(m)}^n(\overline{\Wm})$ and $\Tc_{\xv(m)}^n(\Vm^i)$. If $\yv \in \Tc_{\xv(m)}^n(\overline{\Wm})$ is connected to $\hat{\yv} \in \Tc_{\xv(m)}^n(\Vm^i)$, by the maximality of $\pv_{Y\hat{Y}|X}$ and Lemma \ref{lemmaximal} we have that
		\begin{align}
			\metric(\yv,\xv(m)\big) \leq \metric(\yv,\xv'(r)\big) \ \text{for} \ r = 1,2,\dotsc, a,
		\end{align}
		where $\xv'(r)$ for $r = 1,2,\dotsc,a$ are those that satisfy condition $3$ above. The above inequality implies that if $\xv(m)$ is transmitted and $\yv \in \Tc_{\xv(m)}^n(\overline{\Wm})$ is the channel output, the probability of correct $\metric$-decoding is at most $\frac{1}{a+1}$ because there are $a$ other codewords $\xv'(1),\xv'(2),\dotsc,\xv'(a) \in \Cc_n$ for which the decoding metric is higher, \textit{i.e.} $\metric(\yv,\xv(m)\big) \leq \metric(\yv,\xv'(r)\big)$ for $1 \leq r \leq a$. 
		
		Now we count the number of $\yv\in \Tc_{\xv(m)}^n(\overline{\Wm})$ that cause a $\metric$-decoding error. Recall that from Definition \ref{def:graph1}, the set of all $\yv \in \Tc_{\xv(m)}^n(\overline{\Wm})$ which are connected to a $\hat{\yv} \in \Bc_i$ in graph $\Gc_{\xv(m)}(\pv_{Y\hat{Y}|X})$ was denoted by $\Psi_{21}(\Bc_i)$.  
		In the following, we give a lower bound on $|\Psi_{21}(\Bc_i)|$ based on the Lemma \ref{lemmafhghgh}. So far we have proved the following facts: 
		\begin{enumerate}
			\item There exists a codeword $\xv(m) \in \Cc_n$  and a family of sets $\Fc = \{\Bc_i\}$ such that $\Bc_i \subset \Tc_{\xv(m)}^n(\Vm^i)$ and $|\Bc_i| \geq \frac{a-1}{a}|\Tc_{\xv(m)}^n(\Vm^i)|$.
			\item $\forall \hat\yv \in \Bc_i$ connected to $\yv \in \Tc_{\xv(m)}^n(\overline{\Wm})$ in graph $\Gc_{\xv(m)}(\pv_{Y\hat{Y}|X})$, the following holds, 
			\begin{align}\label{peisbig}
				\PP\big[\hat{m} \neq m\,|\,\yv \text{ \ is recieved},\xv(m) \text{ is sent}\big] \geq \frac{a}{a+1}.	
			\end{align} 
		\end{enumerate}
		We count the number of elements of $\Psi_{21}(\Bc_i)$ in $\Gc_{\xv(m)}(\pv_{Y\hat{Y}|X})$ for  $1 \leq i \leq t$, since the $\metric$-decoder makes errors on elements of $\Psi_{21}(\Bc_i)$. Using the fact that $|\Bc_i| \geq \frac{a-1}{a}|\Tc_{\xv(m)}^n(\Vm^i)|$ and Lemma \ref{lemmafhghgh}  with $\mathcal{V}_1 = \Tc_{\xv(m)}^n(\overline{\Wm})$ and $\mathcal{V}_2 = \Tc_{\xv(m)}^n(\Vm^i)$ we get
		\begin{align}
			\frac{|\Psi_{21}(\Bc_i)|}{|\Tc_{\xv(m)}^n(\overline{\Wm})|} &\geq  \frac{|\Bc_i|}{|\Tc_{\xv(m)}^n(\Vm^i)|} \\ 
			&\geq \frac{\frac{a-1}{a}|\Tc_{\xv(m)}^n(\Vm^i)|}{|\Tc_{\xv(m)}^n(\Vm^i)|}\\
			& = \frac{a-1}{a}.
			\label{tyehewheh}
		\end{align}
		
		\ul{Part 3}
		
		In the remaining part of the proof we relate $\frac{|\Psi_{21}(\Bc_i)|}{|\Tc_{\xv(m)}^n(\overline{\Wm})|}$ to the probability of error. Suppose $\xv(m)$ is sent over the channel and $\yv$ is received.  Note by the definition of conditional type, all elements of $\Tc_{\xv(m)}^n(\overline{\Wm})$ are equally likely to appear at the output of the channel when $\xv(m)$ is sent. Therefore, for every $\yv_0 \in \Tc_{\xv(m)}^n(\overline{\Wm})$,
		\begin{align}
			\PP\big[\yv \in \Tc_{\xv(m)}^n(\overline{\Wm})\ &|\ \xv(m) \text{ \ is sent}\big]\notag\\ &= \sum_{\bar{\yv} \in \Tc_{\xv(m)}^n(\overline{\Wm})}W^n\big(\bar{\yv}|\xv(m)\big)\\
			&= |\Tc_{\xv(m)}^n(\overline{\Wm})| \cdot W^n\big(\yv_0|\xv(m)\big)\label{equiprob}
		\end{align}
		where \eqref{equiprob} follows since $W^n\big(\bar{\yv}|\xv(m)\big)$ is the same for all $\bar{\yv} \in \Tc_{\xv(m)}^n(\overline{\Wm})$.
		Therefore,
		\begin{align}
			&\PP\big[\hat{m} \neq m\,|\, \hat{\pv}_{\yv|\xv(m)} = \overline{\Wm}, \xv(m) \text{ is sent}\big] \\
			&= \PP\big[\hat{m} \neq m \,|\, \yv \in  \Tc_{\xv(m)}^n(\overline{\Wm}), \xv(m) \text{ is sent}\big] \\
			&=\frac{\PP\big[\hat{m} \neq m , \yv \in  \Tc_{\xv(m)}^n(\overline{\Wm})\,|\, \xv(m) \text{ is sent}\big]}{\PP[\yv \in \Tc_{\xv(m)}^n(\overline \Wm)\,|\, \xv(m) \text{ is sent}]}\label{43e1}\\		
			&=\frac{\sum_{\bar{\yv} \in \Tc_{\xv(m)}^n(\overline{\Wm})}\PP\big[\hat{m} \neq m, \yv =\bar{\yv}   \,|\, \xv(m) \text{  is sent}\big]}{\PP\big[\yv \in \Tc_{\xv(m)}^n(\overline \Wm)\,|\, \xv(m) \text{ is sent}\big]}  \\
			&=\frac{1}{\PP\big[\yv \in \Tc_{\xv(m)}^n(\overline\Wm)\,|\, \xv(m) \text{ is sent}\big]}\cdot\notag\\
			&\cdot\sum_{\bar{\yv} \in \Tc_{\xv(m)}^n(\overline{\Wm})}\PP\big[\hat{m} \neq m \,|\,  \yv=\bar{\yv}, \xv(m) \text{  is sent}\big] W^n\big(\bar{\yv}|\xv(m)\big)\\
			%&=\frac{\sum_{\bar{\yv} \in \Tc_{\xv(m)}^n(\overline{\Wm})}\PP\big[\hat{m} \neq m \,|\,  \yv=\bar{\yv}, \xv(m) \text{  is sent}\big]\cdot W^n\big(\bar{\yv}|\xv(m)\big)}{\PP\big[\yv \in \Tc_{\xv(m)}^n(\overline\Wm)\,|\, \xv(m) \text{ is sent}\big]} \\
			&=\frac{1}{|\Tc_{\xv(m)}^n(\overline{\Wm})|} \sum_{\bar{\yv} \in \Tc_{\xv(m)}^n(\overline{\Wm})}\PP\big[\hat{m} \neq m\,|\,\yv=\bar{\yv}, \xv(m) \text{ is sent}\big] \label{43e2} \\
			&\geq \frac{1}{|\Tc_{\xv(m)}^n(\overline{\Wm})|} \sum_{\bar{\yv} \in \Psi_{21}(\Bc_i)}\PP\big[\hat{m} \neq m\,|\,\yv=\bar{\yv},\xv(m) \text{ is sent}\big]  \label{43e3}\\
			&\geq \frac{|\Psi_{21}(\Bc_i)|}{|\Tc_{\xv(m)}^n(\overline{\Wm})|}\frac{a}{a+1} \label{43e4}\\
			&\geq \frac{a-1}{a+1}\label{reitwebq}
		\end{align}
		where \eqref{43e1} follows from definition of conditional probability, \eqref{43e2} follows from \eqref{equiprob}, \eqref{43e3} follows since $\Psi_{21}(\Bc_i)\subseteq\Tc_{\xv(m)}^n(\overline{\Wm})$, \eqref{43e4} follows from \eqref{peisbig} and \eqref{reitwebq} from \eqref{tyehewheh}.
		
		\ul{Part 4}
		
		In the final step, we have the following inequality,
		\begin{align}
			&\PP\big[\hat{m} \neq m\,|\, \hat{\pv}_{\yv |\xv(m)} \in \Nc_{\frac{\epsilon}{2},\pv_X}(W),\xv(m) \text{\ is sent}\big] \notag\\&=\frac{\PP\big[\hat{m} \neq m, \hat{\pv}_{\yv |\xv(m)} \in \Nc_{\frac{\epsilon}{2},\pv_X}(W)\,|\, \xv(m) \text{\ is sent}\big] }{\PP\big[\hat{\pv}_{\yv |\xv(m)} \in \Nc_{\frac{\epsilon}{2},\pv_X}(W)\,|\, \xv(m) \text{\ is sent}\big] }\label{ewt4rw}\\
			&= \sum_{\overline\Wm\in\Nc_{\frac{\epsilon}{2},\pv_X}(W)}  \frac{\PP\big[\hat{m} \neq m , \hat{\pv}_{\yv|\xv(m)} = \overline{\Wm} | \xv(m) \text{\ is sent}\big]}{\PP\big[\hat{\pv}_{\yv|\xv(m)} \in \Nc_{\frac{\epsilon}{2},\pv_X}(W)|\xv(m) \text{\ is sent}\big]} \\
			%&= \sum_{\overline\Wm\in\Nc_{\frac{\epsilon}{2},\pv_X}(\Wm)} \frac{\PP\big[\hat{m} \neq m | \hat{\pv}_{\yv|\xv(m)} = \overline{\Wm} ,\xv(m) \text{\ is sent}\big] \PP\big[\hat{\pv}_{\yv|\xv(m)} = \overline{\Wm}| \xv(m) \text{\ is sent}\big]}{\PP\big[\hat{\pv}_{\yv|\xv(m)} \in \Nc_{\frac{\epsilon}{2},\pv_X}(W) | \xv(m) \text{\ is sent}\big]}  \\
			&= \sum_{\overline\Wm\in\Nc_{\frac{\epsilon}{2},\pv_X}(\Wm)} \frac{\PP\big[\hat{\pv}_{\yv|\xv(m)} = \overline{\Wm}| \xv(m) \text{\ is sent}\big]       }{\PP\big[\hat{\pv}_{\yv|\xv(m)} \in \Nc_{\frac{\epsilon}{2},\pv_X}(W) | \xv(m) \text{\ is sent}\big]}  \cdot\notag\\
			&~~~~~~~~~~~~~~~~~~~\cdot \PP\big[\hat{m} \neq m | \hat{\pv}_{\yv|\xv(m)} = \overline{\Wm} ,\xv(m) \text{\ is sent}\big] \\
			&\geq \frac{a-1}{a+1}\cdot\notag\\
			&~~~\cdot\sum_{\overline\Wm\in\Nc_{\frac{\epsilon}{2},\pv_X}(W)} \frac{\PP\big[\hat{\pv}_{\yv|\xv(m)} = \overline{\Wm} | \xv(m) \text{\ is sent}\big]}{\PP\big[\hat{\pv}_{\yv|\xv(m)} \in \Nc_{\frac{\epsilon}{2},\pv_X}(W) | \xv(m) \text{\ is sent}\big]}\label{fretgir5}  \\
			&= \frac{a-1}{a+1}\label{hthier}
		\end{align}
		where \eqref{ewt4rw} follows from the definition of conditional probability, \eqref{fretgir5}  follows from inequality \eqref{reitwebq}. This concludes the proof.
	\end{IEEEproof}

	%%%%%%%%%%%%%%%%%%%%%%%%%%%%%%
	
	\section{From types to distributions} \label{continuous}
	It is known that if rate $R>0$ is achievable then for any $\epsilon>0$ there exist constant composition codes of rate $R - \epsilon$ whose probability of error tends to $0$. In the following lemma, we prove that if rate $R$ is achievable, then, for any $\epsilon >0$ there exist constant composition codes of rate $R - \epsilon$ with vanishing probability of error that have the additional property that their composition $\pv_n\in\Pc_n(\Xc)$ is such that if  $\pv_n(j) > 0$, then $\pv_n(j) \geq \delta$ for $\delta>0$ independent of $n$, for all $j=1,\dotsc,J$.

	\begin{definition}
		Let $\Cc_n$ be a codebook. We say that $\hat{\Cc}_{\hat{n}}$, for some $\hat n \leq n$, is a $\delta$-reduction of $\Cc_n$ if there exists a sub-codebook $\tilde{\Cc}_n\subseteq \Cc_n$ of composition $\pv_X\in\Pc_n(\Xc)$ such that  $\hat{\Cc}_{\hat{n}}$ is obtained by eliminating all symbols in the set $\Ic = \{j \in \Xc\,|\, \pv_X(j) < \delta\}$ from $\tilde{\Cc}_n$. 
	\end{definition}
	
	\begin{lemma}\label{reduction}
		Let $R>0$ be a rate, then for any $\varepsilon>0$ there exists a $\delta>0$ independent of $n$ such that
		for any codebook $\Cc_n$ of rate $R$ there exists a $\delta$-reduction constant composition codebook $\hat{\Cc}_{\hat{n}}$ with the following properties:
		\begin{align}
			& \hat{n} \geq \big(1 - (J-1)\delta\big)n \label{eq:lemma5eq1}\\
			& P_{e,\rm max}^\metric(\hat{\Cc}_{\hat{n}}) \leq P_{e,\rm max}^\metric(\Cc_n) \label{eq:lemma5eq2}\\
			&\frac{1}{\hat{n}} \log(|\hat{\Cc}_{\hat{n}}|) \geq \frac{1}{n}\log(|\Cc_n|) - \varepsilon + O\left(\frac{\log n}{n}\right)  \label{eq:lemma5eq3}.
		\end{align}	
		
	\end{lemma}
	\begin{IEEEproof}
		For any $n>0$ we know that $|\Pc^n(\Xc)| \leq (n+1)^{J-1}$. Therefore, by the pigeonhole principle, any codebook $\Cc_n$ contains a constant composition sub-codebook $\tilde{\Cc}_n$ of type $\pv_n$  such that $|\tilde\Cc_n|\geq\frac{|\Cc_n|}{(n+1)^{J-1}}$ codewords. Let $\Ic = \{i_1,i_2,...,i_t\}\subset\Xc$ be the set of all symbols $j\in\Xc$ that $\pv_n(j) < \delta$. Then, there are
		\begin{align}
			&\binom{n}{n\pv_n(i_1),n\pv_n(i_2),\dotsc,n\pv_n(i_t)} = \\ &\frac{n!}{\big(n\pv_n(i_1)\big)!\big(n\pv_n(i_2)\big)!\cdots\big(n\pv_n(i_t)\big)!\big(n - \sum_{j = 1}^{t}n\pv_n(i_j)\big)!}
		\end{align}
		possible places for symbols of set $\mathcal{I}$ in a string of length $n$.
		For ease of notation we use the following notation,
		\begin{align}
			\binom{n}{n\pv_n(\mathcal{I})} = \binom{n}{n\pv_n(i_1),n\pv_n(i_2),\dotsc,n\pv_n(i_t)}.
		\end{align}

		As a result, by again using the pigeonhole principle, there exists a sub-codebook $\tilde{\Cc}_n\subseteq\Cc_n$ with $|\tilde{\Cc}_n|\geq\frac{|\Cc_n|}{(n+1)^{J-1}\binom{n}{n\pv_n(\mathcal{I})}}$ codewords where all symbols in set $\mathcal{I}$ are in the same position. By being in the same position we mean that the codewords of $\tilde{\Cc}_n$ have all symbols $i_1, i_2,\dotsc,i_t$ in the same position. Let $\Zc\subset\{1,\dotsc,n\}$ be set of positions where symbols in $\Ic$ are placed. We then form the $\delta$-reducted codebook $\hat{\Cc}_{\hat{n}}$ by shortening the codewords of $\tilde\Cc_n$ such that symbols in positions in $\Zc$ are removed.
		The rate of this codebook is therefore
		\begin{align}
			\frac{1}{\hat{n}}\log(|\hat{\Cc}_{\hat{n}}|) \geq \frac{1}{n}\log(|\hat{\Cc}_{\hat{n}}|) \geq \frac{1}{n}\log\bigg(\frac{|\Cc_n|}{(n+1)^{J-1}\binom{n}{n\pv_n(\mathcal{I})}}\bigg).
		\end{align}
		By using Stirling's factorial formula
		%\beq
		%\sqrt{2\pi} n^{n+\frac 1 2} e^{-n} e^{\frac{1}{12n+1}} < n! < \sqrt{2\pi} n^{n+\frac 1 2} e^{-n} e^{\frac{1}{12n}}
		%\eeq
		we upper-bound the contribution of the multinomial coefficient by the entropy as follows,
		\begin{align}
			&\frac{1}{n}\log\binom{n}{n\pv_n(\mathcal{I})} \\
			&= \frac{1}{n}\log\frac{n!}{\big(n\pv_n(i_1)\big)!\cdots\big(n\pv_n(i_t)\big)!\big(n - \sum_{j = 1}^{t}n\pv_n(i_j)\big)!}
			\\
			%&= \frac{1}{n}\log\frac{n!}{\big(n\pv_n(i_1)\big)!\big(n\pv_n(i_2)\big)!\dotsc\big(n\pv_n(i_t)\big)!\big(n - \sum_{j = 1}^{t}n\pv_n(i_j)\big)!}
			\\
			& \leq H\bigg(\pv_n(i_1),\dotsc,\pv_n(i_t), 1 - \sum_{j = 1}^{t}\pv_n(i_j)\bigg) + O\left(\frac{\log n}{n}\right) \label{hergrr}\\
			&\leq H\big(\underbrace{\delta,\dotsc,\delta}_{J-1}, 1 - (J - 1)\delta\big) +O\left(\frac{\log n}{n}\right)\label{tewre} 
		\end{align}
		where $H(\pi_1,\dotsc,\pi_m) =- \sum_{i=1}^m \pi_i \log \pi_i$ denotes the entropy function of probability mass function with $m$ nonzero mass points with probabilities $\pi_1,\dotsc,\pi_m$ and  \eqref{tewre} follows from observing that $\pv_n(i_1),\pv_n(i_2),..., \pv_n(i_t) \leq \delta$, $t \leq J - 1$ and the fact that $\delta$ can be chosen sufficiently small.
		
		Summarizing, we get the following inequality,
		\begin{align}
			&\frac{1}{\hat{n}}\log(|\hat{\Cc}_{\hat{n}}|) \geq \frac{1}{n} \log(|\hat{\Cc}_{\hat{n}}|) \\
			&\geq R - H\big(\underbrace{\delta,\dotsc,\delta}_{J - 1},1- (J - 1)\delta\big) - (J-1)\frac{\log(n+1)}{n} \\
			&+ O\left(\frac{\log n}{n}\right)
		\end{align}
		Now choosing $\delta$ in a way that $H(\delta,...,\delta,1- (J - 1)\delta) < \varepsilon$  we get the desired result.
		
		It remains to show that
		\begin{align}
			P_{e,\rm max}^\metric(\hat{\Cc}_{\hat{n}}) \leq P_{e,\rm max}^\metric(\Cc_n).
		\end{align}
		This directly follows from the fact that all symbols in $\mathcal{I}$ are in the same position in the codebook $\tilde\Cc_n$. Let us define $\alpha:\mathcal{Y}^n \to \mathcal{Y}^{\hat{n}}$ as the function that takes a string $\yv \in \mathcal{Y}^n$ and  gives $\alpha(\yv) \in\Yc^{\hat n}$ by eliminating the symbols in positions in the set $\mathcal{Z}$. Moreover, let $\mu: \tilde{\Cc}_n \to \hat{\Cc}_{\hat{n}} $ be the function that performs the same operations on the codewords of $\tilde{\Cc}_n$. Then, for any $\yv \in \mathcal{Y}^n$ and $\xv \in \tilde{\Cc}_n$ we have $\metric(\xv, \yv) - \metric\big(\mu(\xv),\alpha(\yv)\big)$ is a function of $\yv$, because all the codewords in $\tilde{\Cc}_n$ have the same symbols at the eliminated entries. As a result, if $\yv$ is decoded to $\xv(\hat{m}) \in \tilde{\Cc}_n$ under $\metric$-decoding, then $\alpha(\yv)$ would be decoded to $\mu(\xv(\hat{m}))$ under $\metric$-decoding. 
		
		Observe that this argument still holds for the case where the metric takes $-\infty$ values. This holds since, whenever $\yv$ is decoded into $\xv(\hat{m}) \in \tilde{\Cc}_n$ under $\metric$-decoding, this necessarily implies that $\metric(\yv,\xv(\hat{m}))$ is finite, which also implies that $\metric(y_i,x_i(\hat{m}))$ are finite for all $1\leq i \leq n$ including indices $i \in \Zc$.  In the case where $\metric(\yv,\xv(\hat{m})) = -\infty$, then this implies that $\metric(\yv,\xv(m)) = -\infty, m=1,\dotsc,M$ and thus, we have a tie, that is decoded as an error.
%Because the sentence "If $\yv$ is decoded to $\xv(\hat{m}) \in \hat{\Cc}_n$ under $\metric$-decoding" implicitly means that $\metric(\yv,\xv(\hat{m}))$ is finite and therefore $\metric(y_i,x_i(\hat{m}))$ is finite for all $1\leq i \leq n$ including indices $i \in \Zc$ Therefore we only have eliminated places that the metric $\metric$ is finite. As a result, the argument still holds.
		 Moreover, since the set $\Zc$ in the lemma has been chosen in such a way that all codewords of $\tilde \Cc_n$ have the same symbols at positions of $\Zc$ we have that the metric between $\yv$ and codewords of $\tilde \Cc_n$ has been finite in the eliminated positions. Let $\xv_{\Zc}$ and $\yv_{\Zc}$ be the symbols of $\xv$ and $\yv$ in positions in set $\Zc$, respectively.
		Now notice that for all ${\yv'} \in \mathcal{Y}^{\hat{n}}$ and all $\xv \in \tilde{\Cc}_n$ 
		\begin{align}\label{grklrg}
			\sum_{\yv \in \alpha^{-1}({\yv}')} W^n(\yv|\xv) &=  \sum_{\yv_{\Zc}}W^{\hat{n}}\big(\yv'|\mu(\xv)\big)W^{n - \hat{n}}(\yv_{\Zc}|\xv_{\Zc}) \\
			&= W^{\hat{n}}\big(\yv'|\mu(\xv)\big)\sum_{\yv_{\Zc}}W^{n - \hat{n}}(\yv_{\Zc}|\xv_{\Zc})\\
			&= W^{\hat{n}}\big(\yv'|\mu(\xv)\big)
		\end{align}
		where \eqref{grklrg} follows from the fact that $W$ is a memoryless channel. Moreover, $\xv_{\Zc}$ and $\yv_{\Zc}$ are strings consisting of symbols of the index set $\Zc$ of $\xv$ and $\yv$ respectively. As a result, the probability of error of any codeword $\xv \in \tilde{\Cc}_n$ is equal to probability of error of $\mu(\xv) \in \hat{\Cc}_{\hat{n}}$. Thus,
		\begin{align}\label{tgteh}
			P_{e,\rm max}^\metric(\hat{\Cc}_{\hat{n}}) = P_{e,\rm max}^\metric(\tilde{\Cc}_n)
		\end{align}
		Since $\tilde{\Cc}_n$ is a sub-codebook of $\Cc_n$, 
		\begin{align}\label{lpepjeje}
			P_{e,\rm max}^\metric(\tilde{\Cc}_n) \leq P_{e,\rm max}^\metric(\Cc_n).
		\end{align}
		Combining \eqref{tgteh} and \eqref{lpepjeje} completes the proof.
	\end{IEEEproof}
	
	The above result is helpful because in order to use the following theorem, we need the frequency of each symbol in any codeword to be proportional to $n$.
	\begin{theorem} [Hoeffding's inequality] \label{Hoef}
		Assume $X_i, i = 1,2,\dotsc,n$ are independent random variables taking values on $[0,1]$. Let $\bar{X} = \frac{1}{n}(X_1 + X_2 + \cdots + X_n)$. Then $\forall \gamma > 0$
		\begin{align}
			\PP\big[|\bar{X} - \mathbb{E}[\bar{X}]\,|\, \geq \gamma\big] \leq e^{-2n{\gamma}^2}.
		\end{align}
	\end{theorem}
	The following lemma shows that the empirical conditional type of the received sequence given the sent message would be close to $\Wm$.
	\begin{lemma} \label{concentrate}
		Let $\xv \in \Tc^n(\pv_X)$ be a codeword, and denote by  $\yv$ the output of channel $W$ when $\xv$ is sent. Then, $\forall \gamma > 0$ we have
		\begin{align}
			\PP\big[\hat{\pv}_{\yv|\xv}(j,k) \in \Nc_{\gamma,\pv_{X}}(W) \,|\, \xv \text{ is sent}\big] >  1 -JK \cdot e ^ {-2n\pv_{\rm min}{\gamma}^2}
		\end{align}
		where $\Nc_{\gamma,\pv_{X}}(W)$ is the channel type neighborhood defined in \eqref{mdakaodapwpr}.
	\end{lemma}
	\begin{IEEEproof}
		Let $(j,k) \in \Xc \times \Yc$ and assume $\pv_{\xv}(j) >0$. We know from the definition of types there are $n\pv_X(j)$ symbols equal to $j\in\Xc$ in $\xv$. Without loss of generality assume, $x_{1} = x_{2} = \cdots = x_{n\pv_X(j)} = j$. Define the random variable $X_i, i = 1,2,\dotsc,n\pv_X(j)$ in the following way,
		\begin{align}
			X_i = \begin{cases} 1 &(y_i,x_i) = (k,j)  \\ 0 & \text{otherwise}.
			\end{cases}
		\end{align}
		As a result, the conditions of Hoeffding's inequality hold for $X_i, i = 1,2,\dotsc,n\pv_X(j)$ and $\EE[X_i] = \PP[X_i = 1] = W(k|j)$. Therefore, we get the following,
		
		\begin{align}\label{vfdjgmiefnaf}
			\PP\big[|W(k|j) - \hat{\pv}_{\yv|\xv}(k|j)| \geq \gamma\,|\,\xv \text{ is sent}\big] \leq e^{-2n\pv_X(j)\gamma^2}.
		\end{align}
		
		As a result, from lower bounding $\pv_X(j)$ by $\pv_{\rm min}$ we get
		\begin{align}\label{liukdytjsrtt}
			\PP\big[|W(k|j) - \hat{\pv}_{\yv|\xv}(k|j)| \geq \gamma\,|\,\xv \text{ is sent}\big] \leq e^{-2n\pv_{\rm min}\gamma^2}.
		\end{align} 
		As a result we have
		
		\begin{align}
			&\PP\big[ \hat{\pv}_{\yv|\xv}(k|j) \in \Nc_{\gamma,\pv_{\xv}}(W)\,|\, \xv \text{ is sent}\big] \notag\\
			&= 1 -  \PP\big[\cup_{j,\pv_{\xv}(j)>0,k}\{|W(k|j) - \hat{\pv}_{\yv|\xv}(k|j)| > \gamma\}\,|\,\xv \text{ is sent}\big]  \\ \label{union}
			&\geq 1 - \sum_{j,\pv_{\xv}(j)>0,k}^{}\PP\big[|W(k|j) - \hat{\pv}_{\yv|\xv}(k|j)| > \gamma\,|\,\xv \text{ is sent}\big]  \\	\label{ho}
			& \geq 1 - \sum_{j,\pv_{\xv}(j)>0,k}^{}e^{-2n\pv_X(j)\gamma^2} \\
			&\geq 1 -JK e ^ {-2n\pv_{\rm min}\gamma^2},
		\end{align}
		where \eqref{union} follows from the union bound, and \eqref{ho} follows from \eqref{vfdjgmiefnaf}. \end{IEEEproof}
	
	The above result shows that when the frequency of every symbol in the codebook grows proportional to $n$, then conditional type of the output string given the sent message will  be close to $W$ with high probability.

	%%%%%%%%%%%%%%%%%%%%%%%%%%%%%%%%%%%
	\section{Proof of the main theorem} \label{sec:end}
	
	In this section, we prove the final part of Theorem \ref{maintheorem} using the material developed in the previous sections. Assume $R = \bar R_\metric(W) + \sigma$ for some $\sigma>0$. 
	Now, choose $\epsilon>0$ small enough such that if $|\overline{V} - V|_{\infty} \leq 2K\epsilon$ for conditional distribution  $V, \overline{V}$, then for any distribution $P_X$ on $\Xc$ we have that
	\begin{align}\label{grerhr}
		&|H(\overline{V}|P_X) - H(V|P_X)| < \frac{\sigma}{4}\\ \label{fdqeeq}
		&|H(\overline{Q}_Y) - H(Q_Y)|<\frac{\sigma}{4}. 
	\end{align}
	where $\overline{Q}_Y,Q_Y$ correspond to output distributions corresponding to input distribution $P_X$ and channel $V,\overline{V}$, respectively. 
	The reader is referred to the Appendix B for a discussion on the choice of $\epsilon$.
	
	From Lemma \ref{reduction} with $\varepsilon = \frac{\sigma}{4}$, for any codebook $\Cc_n$ with $M \geq 2^{nR}$ codewords, there exists a $\delta$-reduction constant composition codebook $\hat{\Cc}_{\hat{n}}$ of length $\hat n$ and type $\hat\pv_{\hat{n}}$ such that  \eqref{eq:lemma5eq1}--\eqref{eq:lemma5eq3} are satisfied. 
	%Now for any codebook $\Cc_n$ with $\left \lfloor 2^{nR} \right \rfloor$ codewords, by using lemma \ref{reduction} by setting $\varepsilon = \frac{\sigma}{4}$ there is a $\delta$-reduction codebook $\hat{\Cc}_{\hat{n}}$ of $\Cc_n$ that has type $\hat\pv_{\hat{n}}$
	%\begin{align}
	%	& \hat{n} \geq \big(1 - (J-1)\delta\big)n \\
	%	&\frac{1}{\hat{n}} \log(|\hat{\Cc}_{\hat{n}}|) \geq \frac{1}{n}\log(|\Cc_n|) - \frac{\sigma}{4} + o_1(n) \\
	%	& P_{e,\rm max}^\metric(\hat{\Cc}_{\hat{n}}) \leq P_{e,\rm max}^\metric(\Cc_n).
	%\end{align}	 
	% 
	Since the required $\delta$ to satisfy the above inequalities is independent of $n$, then choose $N_0$ large enough such that $\epsilon \geq \frac{2K}{N_0 (1-(J-1)\delta )\delta}$. Set $n>N_0$. Choose a maximal joint conditional distribution $P_{Y\hat{Y}|X}$ such that
	%\begin{align}\label{rftw4irj}
	$I(\hat\pv_{\hat{n}},P_{\hat{Y}|X}) \leq  \bar R_\metric(W)$
	%\end{align} 
	%
	and let $V = P_{\hat{Y}|X}$. Such a $P_{Y\hat{Y}|X}$ exists because the set $\Mc_{\rm max}(\metric)\cap \{P_{Y\hat{Y}|X}|P_{Y|X} = W\}$ which is the domain of the minimization in \eqref{eq:upper_bound} is a compact set and the minimizer always exists.
	Moreover, for any conditional distributions  $\widehat{V}$ such that $|\widehat{V} - V|_{\infty} \leq 2K\epsilon$ and $\qv$ being the output distribution corresponding to input type $\hat{\pv}_n$ and channel $V$
	\begin{align} 
		\Big|\max_{\overline{V} \in \Nc_{2K\epsilon,\hat{\pv}_n}(\Vm)}&H(\qv) - H(\widehat{V}|\hat\pv_{\hat{n}})\Big|\notag\\ \label{frief3}
		& \leq |H(\qv) -  H(V|\hat\pv_{\hat{n}})| + \frac{\sigma}{2}\\
		&= I(\hat\pv_{\hat{n}},V)+\frac{\sigma}{2} \label{dqrr23}
	\end{align}
	where \eqref{frief3} follows from \eqref{grerhr} and \eqref{fdqeeq}.

	Suppose $\Nc_{\epsilon,{\hat\pv}_{\hat n}}(V) = \{\Vm^1,\Vm^2,\dotsc,\Vm^t\}$ and $\qv^i$ be the output type corresponding to input type $\hat\pv_{\hat{n}}$ and conditional type $\Vm^i$. For any $1 \leq i \leq t$ we have
	\begin{align}
		\frac{1}{\hat{n}}\log\frac{\max_{1\leq s \leq t}|\Tc^{\hat{n}}(\qv^s)|}{|\Tc^{\hat{n}}_{\xv}(\Vm^i)|} &= \frac{1}{\hat{n}}\log\frac{2^{\hat{n}\left(H(\qv^{i'})+O\left(\frac{\log \hat{n}}{\hat{n}}\right)\right)}}{2^{\hat{n}\left(H(V^i|\hat\pv_{\hat{n}})+O\left(\frac{\log \hat{n}}{\hat{n}}\right)\right)}}\\
		%& = H(\qv^{i'}) - H(V^i|\hat\pv_{\hat{n}}) + O\left(\frac{\log \hat{n}}{\hat{n}}\right) \\ \label{re4t}
		&\leq I(\hat\pv_{\hat{n}},V) + \frac{\sigma}{2}+ O\left(\frac{\log \hat{n}}{\hat{n}}\right)\label{re4t}
	\end{align}
	where $i' = \argmax_{1\leq s \leq t}|\Tc^{\hat{n}}(\qv^s)|$ and \eqref{re4t} follows form \eqref{dqrr23} (see \cite[Ch. 2]{csiszar2011information}  for details about the $\frac{\log \hat{n}}{\hat{n}}$ terms.) 
	Now, for $n>N_0$  we have from \eqref{eq:lemma5eq3} with $\varepsilon = \frac{\sigma}{4}$, \eqref{re4t}  and the condition $I(\hat\pv_{\hat{n}},P_{\hat{Y}|X}) \leq  \bar R_\metric(W)$ that
	\begin{align}
		|\hat \Cc_{\hat n}|\frac{|\Tc^{\hat{n}}_{\xv}(\Vm^i)|}{\max_{1\leq s \leq t}|\Tc^{\hat{n}}(\qv^s)|} \geq 2^{\hat{n}\left(R - \frac{\sigma}{4}-I(\hat\pv_{\hat{n}},V) - \frac{\sigma}{2} + O\left(\frac{\log \hat{n}}{\hat{n}}\right)\right)}.
	\end{align}
	As a result,
	\begin{align}
		|\hat \Cc_{\hat n}| |\Tc^{\hat{n}}_{\xv}(\Vm^i)| \geq 2^{\hat{n}\left(\frac{\sigma}{4}+O\left(\frac{\log \hat{n}}{\hat{n}}\right)\right)} \max_{1\leq s \leq t}|\Tc^{\hat{n}}(\qv^s)|.
	\end{align}

	Setting $a = \Big \lfloor \frac{2^{\frac{1}{2}\hat n\left(\frac{\sigma}{4}+O\left(\frac{\log \hat{n}}{\hat{n}}\right)\right)}}{(\hat n+1)^{J(K-1)}} \Big \rfloor$ validates the conditions of Theorem \ref{theo2}. As a result, there exists $\xv(m) \in \hat\Cc_{\hat n}$ such that
	\begin{align}
		\PP\Big[\hat{m} \neq m\,\big|\, \hat{\pv}_{\yv|{\xv(m)}} \in \Nc_{\frac{\epsilon}{2},\hat\pv_{\hat{n}}}(W),{\xv(m)} \text{\ is sent}\Big] > 1 - \frac{2}{a+1}.
	\end{align}
	According to the definition of limit, choosing $N_1$ such that if $n>N_1$ is large enough, we can bound
	\begin{align}
		a %&= \Big \lfloor \frac{2^{\hat n\frac{1}{2}\left(\frac{\sigma}{4}+O\left(\frac{\log \hat{n}}{\hat{n}}\right)\right)}}{(\hat n+1)^{J(K-1)}} \Big \rfloor \\
		&> \frac{1}{2}\cdot  \frac{2^{\frac{1}{2}\hat n\left(\frac{\sigma}{4}+O\left(\frac{\log \hat{n}}{\hat{n}}\right)\right)}}{(\hat n+1)^{J(K-1)}}\\
		& \geq 2^{\frac{1}{2}\hat{n}\left(\frac{\sigma}{4} +O\left(\frac{\log \hat{n}}{\hat{n}}\right) - 2J(K - 1)\frac{\log(\hat{n}+1)}{\hat{n}} - \frac{\log(2)}{\hat{n}}\right)}.
	\end{align}
	Finally, we write
	\begin{align}
		&\hspace{-2mm}P_{e,\rm max}^\metric(\Cc_n) \geq P_{e,\rm max}^\metric(\hat{\Cc}_{\hat n}) \notag\\
		&= \max_{m\in\{1,\dotsc,M\}}\PP\big[\hat{m} \neq m\,|\, \xv(m) \text{\ is sent}\big] \\
		&\geq \max_{m\in\{1,\dotsc,M\}} \PP\big[\hat{m} \neq m\,|\, \hat{\pv}_{\yv|\xv(m)} \in \Nc_{\frac{\epsilon}{2},\hat\pv_{\hat{n}}}(W), \xv(m) \text{\ is sent}\big] \notag\\
		&~~~~\cdot\PP\big[\hat{\pv}_{\yv|\xv(m)} \in \Nc_{\frac{\epsilon}{2},\hat\pv_{\hat{n}}}(W)| \xv(m) \text{\ is sent}\big] \label{eq:almostend}\\
		&\geq \Big(1-\frac{2}{a+1}\Big)\Big(1 - JK2^{-2\hat n\delta\frac{\epsilon ^2}{4}}\Big) \label{eq:almostendend}\\
		&\geq 1 - 2^{-\hat{n}\bar E_\metric(R)}
	\end{align}
	where $\xv(m)$ is the codeword sent from codebook $\hat \Cc_{\hat n}$, \eqref{eq:almostendend} follows from applying Theorem \ref{theo2} to the first probability in \eqref{eq:almostend} and Lemma \ref{concentrate} to the second probability in \eqref{eq:almostend}, 
	where $\bar E_\metric(R) \eqdef \min\left\{\frac{\delta\epsilon^2}{2} - \log\frac{JK}{\hat n},\frac{1}{2}\left(\frac{\sigma}{4}+O\left(\frac{\log \hat{n}}{\hat{n}}\right)\right)\right\}$. Setting $n$ larger than $\max\{N_0,N_1\}$ yields the desired result.
	%\begin{remark}
	%By the same method we can show that the probability of having a type conflict error over channel $W$ goes to $1$ exponentially above the capacity. As a result if an oracle tells the ML decoder the exact conditional type of received $\yv$ and sent codeword $\xv$ \textit{i.e.} $\hat{\pv}_{\yv|\xv}$, still the probability of error tends to $1$ exponentially. 
	%\end{remark}

	%%%%%%%%%%%%%%%%%%%%%%%%%%%%%%%%%%
	\section{Convexity Analysis}\label{convexity}
	In this section, we show that the optimization \eqref{eq:upper_bound} is a convex-concave saddlepoint problem. First we argue that the constraints induce a convex set.
	\begin{lemma}
		For any channel $W$ and metric $\metric$, the set of joint conditional distributions $P_{Y\hat{Y}|X}$ satisfying both $P_{Y\hat{Y}|X} \in \Mc_{\rm max}(\metric)$ and  $P_{Y|X} = W$, is a convex set.
	\end{lemma}
	\begin{IEEEproof}
		Let $P_{Y\hat{Y}|X}$ and $P'_{Y\hat{Y}|X}$ both satisfy the above constraints. Therefore, for any $0< \lambda<1 $ we have
		\begin{align}
			\lambda P_{Y|X}+(1-\lambda)P'_{Y|X} = W.
		\end{align}
		In addition, if for some $k_1,k_2$ we have $j \notin \Sc_\metric(k_1,k_2)$, both $P_{Y\hat{Y}|X}(k_1,k_2|j)$ and $P'_{Y\hat{Y}|X}(k_1,k_2|j)$ are equal to zero, and so is any linear combination of them. Therefore,
		\begin{align}
			\lambda P_{Y\hat{Y}|X} + (1-\lambda)P'_{Y\hat{Y}|X} \in \Mc_{\rm max}(\metric).
		\end{align}
	\end{IEEEproof}
	
	Moreover, $I(P_X,P_{\hat{Y}|X})$ is convex in terms of $P_{\hat{Y}|X}$, and concave in terms of $P_X$. Since $P_{\hat Y|X}$ is a linear function of $P_{Y\hat Y|X}$, we get that $I(P_X,P_{\hat Y|X})$ is also convex in terms of $P_{Y\hat Y|X}$. Therefore from the minimax theorem \cite{minimax},
	\begin{align}
		\bar R_\metric(W) &=\max_{P_X}\min_{\substack{P_{Y\hat{Y}|X} \in \Mc_{\rm max}(\metric)\\ P_{Y|X} = W}} I(P_X,P_{\hat{Y}|X})\\ \label{ferf3r3}
		&= \min_{\substack{P_{Y\hat{Y}|X} \in \Mc_{\rm max}(\metric)\\ P_{Y|X} = W}}\max_{P_X} I(P_X,P_{\hat{Y}|X}) \\
		&= \min_{\substack{P_{Y\hat{Y}|X} \in \Mc_{\rm max}(\metric)\\ P_{Y|X} = W}} C(P_{\hat{Y}|X}).
	\end{align}

	The rest of this section is devoted to deriving the KKT conditions for the optimization problem in \eqref{eq:upper_bound}. Given that $I(P_X,P_{\hat{Y}|X})$ is convex in $P_{Y\hat{Y}|X}$ and concave in $P_X$, then the KKT conditions are sufficient for global optimality \cite{boyd}.
	For convenience, we define $f(P_X,P_{Y\hat{Y}|X})\triangleq I(P_X,P_{\hat{Y}|X})$ and rewrite the optimization problem in \eqref{eq:upper_bound} as,
	\begin{align}\label{gtpfkemif}
		\bar R_\metric(W) = \max_{P_X}\min_{\substack{P_{Y\hat{Y}|X} \in \Mc_{\rm max}(\metric)\\ P_{Y|X} = W}} f(P_X,P_{Y\hat{Y}|X}).
	\end{align}
	Let $P_X^*,P^*_{Y\hat Y|X}$ be the optimal input and joint conditional distributions in \eqref{gtpfkemif} and $Q^*_{\hat Y}$ be the output distribution induced by $P_X^*$ and $P^*_{\hat Y|X}$. Then for $P_X^*$ we have the following constraints:
	\begin{align} \label{ofeiwruer}
		& P_X^*(j) \geq 0, ~~\forall j \in \Xc\\ \label{pderitgr4t}
		&\sum_{j \in \Xc}^{} P_X^*(j) = 1.
	\end{align}
	Let $\mu_j, j = 1,2,\dotsc,J$ be the Lagrange multipliers corresponding the inequalities in \eqref{ofeiwruer} and $\rho$ be the Lagrange multiplier corresponding to \eqref{pderitgr4t}. Therefore, from stationarity we have
	\begin{align} \label{gvkirnfuwe}
		\frac{\partial}{\partial P_X(j)}f(P_X,P^*_{Y\hat Y|X}) \bigg|_{P_X= P_X^*} = \rho + \mu_j.
	\end{align}
	Then from the complementary slackness  we have $\mu_j\, P_X^*(j) = 0$ and from the dual feasibility we have $\mu_j \geq 0$ which leads to the separation of the equations \eqref{gvkirnfuwe} into two cases \cite{boyd}. 
	If $P_X^*(j) > 0$ 
	\begin{align} \label{mrgmirt}
		\frac{\partial}{\partial P_X(j)}f(P_X,P^*_{Y\hat Y|X}) \bigg|_{P_X= P_X^*} = \rho.
	\end{align}
	And when $P_X^*(j) = 0$ we have
	\begin{align} \label{ghpfeir}
		\frac{\partial}{\partial P_X(j)}f(P_X,P^*_{Y\hat Y|X})\bigg|_{P_X = P_X^*} \leq \rho.
	\end{align}
	Note that, because there is no other constraint on $\mu_j$, all of the KKT conditions are summarized  in \eqref{mrgmirt} and \eqref{ghpfeir}.
	Moreover, computing the derivatives in \eqref{mrgmirt} and \eqref{ghpfeir} gives 
	\begin{align}\label{derivative1}
		\frac{\partial}{\partial P_X(j)}f(P_X,&P^*_{Y\hat Y|X})\bigg|_{P_X = P_X^*}\notag\\
		& =  \sum_{k \in \Yc}^{}P^*_{\hat Y|X}(k|j)\log \frac{P^*_{\hat Y|X}(k|j)}{Q^*_{\hat Y}(k)} -1.
	\end{align}
	
	Similarly, for $P^*_{Y\hat Y|X}$ we have the following constraints.
	For all $j,k_1,k_2 \in \Xc \times \Yc \times \Yc$,
	\begin{align} \label{erfjewhuherrh}
		&P^*_{Y\hat Y|X}(k_1,k_2|j) \geq 0, \\ \label{rfnerenqw}
		&P^*_{Y\hat Y|X}(k_1,k_2|j) = 0, \,\text{if} \ j \notin \Sc_\metric(k_1,k_2)
	\end{align}
	where \eqref{erfjewhuherrh} corresponds to $P^*_{Y\hat Y|X}(k_1,k_2|j)$ being a distribution and \eqref{rfnerenqw} corresponds to $P^*_{Y\hat Y|X}(k_1,k_2|j) \in \Mc_{\rm max}(\metric)$. Moreover from the constraint $P_{Y|X} = W$ we get for all $j,k_1 \in \Xc \times \Yc$
	\begin{align} \label{ropefgertfewur}
		\sum_{k_2}^{}P^*_{Y\hat Y|X}(k_1,k_2|j) = W(k_1|j).
	\end{align}
	For the ease of notation, we skip the step of explicitly considering a Lagrange multiplier for \eqref{erfjewhuherrh}. Details follow similarly to the above derivation. If we use a Lagrange multiplier $\lambda_{j,k_1}$ for each of the conditions in \eqref{ropefgertfewur}, we have when $P^*_{Y\hat{Y}|X}(k_1,k_2|j) > 0$ 
	\begin{align}
		\frac{\partial}{\partial P_{Y\hat{Y}|X}(k_1,k_2|j)}f(P_X^*,P_{Y\hat Y|X}) \bigg|_{P_{Y\hat Y|X} = P^*_{Y\hat Y|X}} = \lambda_{j,k_1}
	\end{align}
	and when $P^*_{Y\hat{Y}|X}(k_1,k_2|j) = 0$ and $j \in \Sc_\metric(k_1,k_2)$ we should have
	\begin{align}
		\frac{\partial}{\partial P_{Y\hat{Y}|X}(k_1,k_2|j)}f(P_X^*,P_{Y\hat Y|X})\bigg|_{P_{Y\hat Y|X} = P^*_{Y\hat Y|X}} \geq \lambda_{j,k_1}.
	\end{align}
	Explicitly computing the derivative gives
	\begin{align}
		\frac{\partial}{\partial P_{Y\hat{Y}|X}(k_1,k_2|j)} &f(P_X^*,P_{Y\hat Y|X})\bigg|_{P_{Y\hat Y|X} = 
		 P^*_{Y\hat Y|X}} \notag\\
	 &= P_X^*(j) \log \frac{P^*_{\hat Y|X}(k_2|j)}{Q^*_{\hat Y}(k_2)}.
		\label{eq:partial_single}
	\end{align}
	
	Summarizing, for the KKT optimality conditions of \eqref{gtpfkemif} we get the following inequalities,
	\begin{enumerate}
		\item If $P_X^*(j) > 0$,
		\begin{align}
			\sum_{k \in \Yc}^{}P^*_{\hat Y|X}(k|j)\log \frac{P^*_{\hat Y|X}(k|j)}{Q^*_{\hat Y}(k)} = 1+\rho.
		\end{align}
		\item If  $P_X^*(j) = 0$,
		\begin{align}
			\sum_{k \in \Yc}^{}P^*_{\hat Y|X}(k|j)\log \frac{P^*_{\hat Y|X}(k|j)}{Q^*_{\hat Y}(k)} \leq 1+ \rho.
		\end{align}
		
		\item If $P^*_{Y\hat{Y}|X}(k_1,k_2|j) > 0$,
		\begin{align} \label{firstoptimal}
			P_X^*(j) \log \frac{P^*_{\hat Y|X}(k_2|j)}{Q^*_{\hat Y}(k_2)} = \lambda_{j,k_1}.
		\end{align}
		\item If $P^*_{Y\hat{Y}|X}(k_1,k_2|j) = 0$ and $j \in \Sc_\metric(k_1,k_2)$,
		\begin{align}\label{secondoptimal}
			P_X^*(j) \log \frac{P^*_{\hat Y|X}(k_2|j)}{Q^*_{\hat Y}(k_2)} \geq \lambda_{j,k_1}.
		\end{align}
	\end{enumerate}
	
	In the next sections, we employ the above KKT conditions to efficiently compute $\bar R_\metric(W)$ and to analyze the multiletter version of the bound.
	
	%%%%%%%%%%%%%%%%%%%%%%%%%%%%%%%%%%%%%%%%%%%
	\section{Computation of $\bar R_\metric(W)$} \label{sec:comp}
	
	In this section, we turn to the computation of the proposed upper bound $\bar R_\metric(W)$. Before describing the algorithm in detail, we introduce a number of concepts related to convex-concave optimization. Let $\Dc \subset \RR^n$ be an open convex set. The standard inner product on $\RR^n$ is denoted by $\langle \cdot,\cdot \rangle$. A mirror map is a function $\Psi: \Dc \to \RR$ with the following properties:
	%Function $\Psi: \Dc \to \RR$ is called a mirror map if it has the following properties.
	\begin{enumerate}
		\item $\Psi$ is strictly convex and continuously differentiable on $\Dc$, where strict convexity means that for all $\vv_1,\vv_2 \in \Dc$
		\begin{align}
			\Psi(\vv_1) - \Psi(\vv_2) - \big\langle \nabla \Psi(\vv_2), \vv_1 - \vv_2 \big\rangle > 0,
		\end{align}
		\item The range of $\nabla \Psi$ is all of $\RR^n$ \textit{i.e.} $\nabla\Psi(\Dc) = \RR^n$,
		
		\item The gradient of $\Psi$ diverges on the boundary of $\Dc$, denoted by $\partial \Dc$, that is
		\begin{align}
			\lim_{\vv \to \partial \Dc} \lVert \nabla \Psi(\vv) \rVert = \infty,
		\end{align}
		where $\lVert \cdot \rVert$ denotes the Euclidean norm.
		%Where $\partial \Dc$ is the boundary of $\Dc$.
	\end{enumerate}
	
	The Bregman divergence $B_{\Psi}(\cdot,\cdot): \Dc \times \Dc \to \RR$ with respect to a mirror map $\Psi$ is defined as
	\begin{align}
		B_{\Psi}(\vv_1,\vv_2) = \Psi(\vv_1) - \Psi(\vv_2) - \big\langle \nabla \Psi(\vv_2), \vv_1 - \vv_2 \big\rangle.
	\end{align}
	Let $\Dc \subset \RR^n$ be a convex set. Function $h:\Dc \to \RR$ is said to be $\alpha$-strongly convex with respect to norm $|\cdot|$ if it is differentiable on $\Dc$ and for all $\vv_1,\vv_2 \in \Dc$ we have 
	\begin{align} \label{strongconvex}
		h(\vv_1) - h(\vv_2) - \big\langle \nabla h(\vv_2) , \vv_1 - \vv_2 \big\rangle \geq \frac{\alpha}{2}|\vv_1-\vv_2|^2,
	\end{align} 
	where the norm $|\cdot|$ is not necessarily induced by the standard inner product, \textit{i.e.} it is not necessarily the Euclidean norm.
	If the mirror map $\Psi : \Dc \to \RR$ is $1$-strongly convex with respect to the norm $|\cdot|$ then from the definition \eqref{strongconvex} for all $\vv_1,\vv_2 \in \Dc$ we have
	\begin{align}\label{oxwuewnrnr}
		B_{\Psi}(\vv_1,\vv_2) \geq \frac{1}{2}|\vv_1-\vv_2|^2.
	\end{align}
	
	We aim to compute the value of the following saddlepoint problem,
	\begin{align}
		\bar R_\metric(W) = \max_{P_X}\min_{\substack{P_{Y\hat{Y}|X} \in \Mc_{\rm max}(\metric)\\ P_{Y|X} = W}} f(P_X,P_{Y\hat{Y}|X}).
		\label{eq:lkjhlkjga}
	\end{align}
	
	For ease of notation and consistency we define $\Ec_1$ and $\Ec_2$ be the constraint sets corresponding to the maximization and minimization, respectively,  
	\begin{align}
		&\Ec_1 = \big\{\vv \in \RR^J ~|~ v(j) \geq 0, \sum_{j=1}^{J} v(j) = 1\big\} \\
		&\Ec_2 = \big\{\uv \in \RR^{J\times K \times K} ~\big|~ \sum_{k_2=1}^{K} u(j,k_1,k_2) = W(k_1|j), \notag \\
		& ~~u(j,k_1,k_2) \geq 0,\, u(j,k_1,k_2) = 0 \text{ if } j \notin \Sc_\metric(k_1,k_2) \big\}
	\end{align}
	where $\Ec_1$ corresponds to the set of distributions over $\Xc$ and $\Ec_2$ corresponds to the set of maximal joint conditional distributions $P_{Y\hat Y|X}$ such that $P_{Y|X} = W$ \textit{i.e.} $\Mc_{\rm max}(\metric) \cap  \{P_{Y\hat{Y}|X}|P_{Y|X} = W \}$.  There is a natural bijection between the two sets by mapping $\uv$ to $P_{Y\hat Y|X}$ such that for every $(j,k_1,k_2) \in \Xc \times \Yc \times \Yc$ we have $\uv(j,k_1,k_2) = P_{Y\hat Y|X}(k_1,k_2|j)$. With a slight abuse of notation let $f$ be defined for vectors $\vv \in \Ec_1, \uv \in \Ec_2$ as it is defined for their corresponding distributions $P_X, P_{Y\hat Y|X}$ in the previous section, \textit{i.e.}, $f(\vv,\uv) \triangleq f(P_X,P_{Y \hat Y|X})$. Therefore, with a slight abuse of notation, we rewrite the saddlepoint problem \eqref{eq:lkjhlkjga} as
	\begin{align}
		\bar R_\metric(W) = \max_{\vv \in \Ec_1}\min_{\uv \in \Ec_2} f(\vv,\uv).
		\label{eq:ewlrskjhk}
	\end{align}
	In the rest of this section, whenever $\uv$ is used, it is considered that $u(j,k_1,k_2) = 0$ if $j \notin \Sc_\metric(k_1,k_2)$, i.e., that the corresponding $P_{Y\hat Y|X}\in \Mc_{\rm max}(\metric)$.
	Additionally, we choose 
	\beq
	\Dc_1 = \{\vv \in \RR^J | 0 \leq \vv(j) , 0 \leq j \leq J\}
	\eeq
	and 
	\begin{align}
		\Dc_2 = \{\uv &\in \RR^{J \times K \times K} | 0 \leq y(j,k_1,k_2), 0\leq j \leq J, \notag\\
		&0 \leq k_1,k_2 \leq K , u(j,k_1,k_2) = 0 \textit{\ if\ } j \notin \Sc_\metric(k_1,k_2) \}.
	\end{align}
	
	It is known that the function $\Psi_1(\vv) = \sum_{i} v(i) \log v(i)$ is a $1$-strongly convex mirror map on $\Dc_1$ with respect to norm $|\cdot|_1$ \cite{Bubeck}. Additionally, let $\Psi_2(\uv) = \sum_{j,k_1,k_2} \indicator\{j \in \Sc_\metric(k_1,k_2) \} u(j,k_1,k_2)\log u(j,k_1,k_2)$. Note that $\nabla \Psi_2$ is surjective on
	$\big\{\uv \in \RR^{J\times K \times K} ~\big| u(j,k_1,k_2) = 0 \text{ if } j \notin \Sc_\metric(k_1,k_2) \big\}$. Moreover, in all of the computations regarding $\uv$ we only use the entries $u(j,k_1,k_2)$ such that $j \in \Sc_\metric(k_1,k_2)$ and ignore all other entries, \textit{i.e.}, they are set to $0$ from the beginning of the algorithm and never change. Therefore, with a slight abuse of notation we say $\Psi_2$ is a $1$-strongly convex mirror map on $\Dc_2$ with respect to norm $|\cdot|_1$. Note that for $\Psi_2$ being a mirror map, from the definition we need it to be surjective on $\RR^{J \times K \times K}$, but since in the whole algorithm only the coordinates $(j,k_1,k_2)$ are used such that $j \in \Sc_\metric(k_1,k_2)$ and $\nabla \Psi_2$ is surjective on $\big\{\uv \in \RR^{J\times K \times K} ~\big| u(j,k_1,k_2) = 0 \text{ if } j \notin \Sc_\metric(k_1,k_2) \big\}$ all the properties of a mirror map are preserved. Moreover, the corresponding Bregman divergences $B_{\Psi_1}$ and $B_{\Psi_2}$ are given by
	\begin{align} \label{cmiefueafra}
		&B_{\Psi_1}(\vv_1,\vv_2) = \sum_{i} v_1(i) \log \frac{v_1(i)}{v_2(i)} - v_1(i) + v_2(i) \\ \label{ewrhw7rwer}
		&B_{\Psi_2}(\uv_1,\uv_2) \notag\\
		&= \sum_{j,k_1,k_2} \indicator\{j \in \Sc_\metric(k_1,k_2)\} \bigg[u_1(j,k_1,k_2) \log \frac{u_1(j,k_1,k_2)}{u_2(j,k_1,k_2)} \notag\\
		&~~~~~- u_1(j,k_1,k_2) + u_2(j,k_1,k_2)\bigg].
	\end{align}
	Note that when $\vv_1,\vv_2 \in \Ec_1$ the Bregman divergence $B_{\Psi_1}(\vv_1,\vv_2)$ reduces to relative entropy between $\vv_1$ and $\vv_2$.
	
	It is known that the Bregman divergence \eqref{cmiefueafra} is jointly convex in its arguments \cite{Bubeck}. Similarly, \eqref{ewrhw7rwer} is  jointly convex in its arguments as well.

	We will use the algorithm mirror prox \cite{Nemirovski}, known to be able to iteratively find the saddlepoint for convex-concave optimization problems where the gradients $\nabla_\vv f(\vv,\uv)$ and $\nabla_\uv f(\vv,\uv)$ are Lipshitz functions. Unfortunately, this condition does not hold in our problem, because the gradient is not necessarily finite on the boundries of both $\Ec_1,\Ec_2$. Therefore, we need the following result to control the growth of the gradient. Then using this fact, we add an additional step to the standard mirror prox algorithm and show that it converges to the saddlepoint.
	Note that the notation $\nabla_{\vv = \vv_0} f(\vv, \uv_0)$ represents the gradient of $f(\vv,\uv_0)$ at point $\vv_0$; $\nabla_{\uv = \uv_0} f(\vv_0, \uv)$ is defined accordingly.
	\begin{lemma}\label{bounded gradient}
		Let $\vv_0, \uv_0$ be defined as
		\begin{align}\label{x0def}
			\vv_0(j) = \frac{1}{J}, \ \forall j \in \Xc 
		\end{align}
		for all $(j,k_1,k_2) \in \Xc \times \Yc \times \Yc.$
		\begin{align}
			 \label{y0def}
			\uv_0(j,k_1,k_2) = \frac{W(k_1|j)\indicator\{j \in \Sc_\metric(k_1,k_2)\})}{|\sum_{k_2} \indicator\{j \in \Sc_\metric(k_1,k_2)\} |}.
		\end{align}
		Let $\kappa = \frac{1}{T}$, then for any $(\vv',\uv') \in \Ec_1 \times \Ec_2$
		\begin{align} \label{mcxqaiwenqw}
			&|\nabla_{\vv = (1-\kappa)\vv'+\kappa \vv_0} f(\vv,(1-\kappa)\uv'+\kappa \uv_0)|_{\infty} \notag\\
			& ~~~\leq  \log(K) + \log \frac{TJ}{W_{\rm min}} + 1 \\ \label{dvtawvdgvcavsf}
			&|\nabla_{\uv = (1-\kappa)\uv'+\kappa \uv_0} f((1-\kappa)\vv'+\kappa \vv_0,\uv) |_{\infty} \notag\\
			&~~~ \leq  \log \frac{TK}{W_{\rm min}} + \log \frac{TJ}{W_{\rm min}},
		\end{align}
		where $W_{\rm min} = \min_{\substack{j \in \Xc,k \in \Yc:\\W(k|j)> 0}} W(k|j)$.
	\end{lemma}
	\begin{IEEEproof}
		In the following expressions $P_X,P_{Y \hat{Y}|X}$ correspond to $(1-\kappa)\vv'+\kappa \vv_0 , (1-\kappa)\uv'+\kappa \uv_0$, respectively. Note that every entry of $(1-\kappa)\vv'+\kappa \vv_0$ is greater than or equal to $\frac{1}{TJ}$. As a result, every entry of $Q_{\hat Y}$, which is output distribution corresponding to $P_X,P_{Y \hat{Y}|X}$, is either $0$ or greater than or equal to $\frac{W_{\rm min}}{TJ}$. Recall that the $j$-th entry of $|\nabla_{\vv = (1-\kappa)\vv'+\kappa \vv_0} f(\vv,(1-\kappa)\uv'+\kappa \uv_0)|_{\infty}$ is equal to $\frac{\partial}{\partial P_X(j)}f(P_X,P_{Y\hat Y|X})$.
		Therefore, \eqref{mcxqaiwenqw} follows by,
		\begin{align}
			&\left|\frac{\partial}{\partial P_X(j)}f(P_X,P_{Y\hat Y|X}) \right| \notag\\
			&=  \left| \sum_{k \in \Yc}^{}P_{\hat Y|X}(k|j)\log \frac{P_{\hat Y|X}(k|j)}{Q_{\hat Y}(k)} -1 \right| \\
			&= \left| -H(\hat Y |X = j) - \sum_{k \in \Yc} P_{\hat Y|X}(k|j) \log(Q_{\hat Y}(k)) -1 \right|  \\
			&\leq \log(K) + \log \frac{TJ}{W_{\rm min}} + 1.
		\end{align}
		Recall that the entries of $|\nabla_{\vv = (1-\kappa)\vv'+\kappa \vv_0} f(\vv,(1-\kappa)\uv'+\kappa \uv_0)|_{\infty}$ are equal to $\frac{\partial}{\partial P_{Y\hat{Y}|X}(k_1,k_2|j)}f(P_X,P_{Y\hat Y|X})$ for $1 \leq k_1,k_2 \leq K$. Moreover, when $j \in \Sc_\metric(k_1,k_2)$ then, $P_{\hat Y |X}(k_2|j) \geq P_{Y \hat Y |X}(k_1, k_2|j) \geq \frac{W_{\rm min}}{TK}$. As a result \eqref{dvtawvdgvcavsf} follows from,
		\begin{align}
			&\left|\frac{\partial}{\partial P_{Y\hat{Y}|X}(k_1,k_2|j)}f(P_X,P_{Y\hat Y|X}) \right|\notag\\
			& = \left|P_X(j) \log \frac{P_{\hat Y|X}(k_2|j)}{Q_{\hat Y}(k_2)}\right| \\
			&\leq \left|\log P_{\hat Y|X}(k_2|j)\right| + \left|\log Q_{\hat Y}(k_2) \right|\\
			&\leq  \log \frac{TK}{W_{\rm min}} + \log \frac{TJ}{W_{\rm min}} 
		\end{align}
	\end{IEEEproof}
	
	For ease of notation, let 
	\begin{align}
		G = \max \bigg\{ \log(K) + \log \frac{TJ}{W_{\rm min}} + 1 ,\log \frac{TK}{W_{\rm min}} + \log \frac{TJ}{W_{\rm min}}\bigg\}
	\end{align} in the rest of the section.
	From the choices of $\vv_0,\uv_0$ in \eqref{x0def} and \eqref{y0def}, and \eqref{cmiefueafra} and \eqref{ewrhw7rwer} we get
	\begin{align} \label{gmritesur}
		&\max_{\vv \in \Ec_1} B_{\Psi_1}(\vv,\vv_0) \leq \log(J) \\ \label{cmeherge}
		&\max_{\uv \in \Ec_2} B_{\Psi_2}(\uv,\uv_0) \leq  J\log \frac{K}{W_{\rm min}}.
	\end{align}
	Here \eqref{gmritesur} holds since the relative entropy between a distribution and the uniform distribution is bounded by the logarithm of the alphabet cardinality. Furthermore, from definition \eqref{y0def} all of the entries of $\uv_0$ are either $0$ or not less than $\frac{W_{\rm min}}{K}$. Additionally, by definition of set $\Ec_2$, $\uv$ is equal to zero at entries that $\uv_0$ equals to zero. Using these two facts \eqref{cmeherge} follows.
	Let $\vv_t$ and $\uv_t$, $t = 1,2,\dotsc, T$ be defined by the following iterative algorithm, where $T$ is the maximum number of iterations. The computation $\vv_{t}$ is described in Algorithm \ref{cbayfvyttyt}, where $\eta_t$ is the stepsize at iteration $t$. From the definition of mirror map, the range of  $\nabla \Psi_1$ is $R^J$, guaranteeing the existence of  $\tilde{\vv} _{t}$ in the gradient step of the above algorithm. Similarly, the computation of $\uv_{t}$ is described in Algorithm \ref{hprawihte}.

	\begin{algorithm}[t]
		\SetAlgoLined
		Initialize:  choose $\vv_0,\uv_0$ from \eqref{x0def} and \eqref{y0def}, respectively \\
		\For{$t = 1,2,\dotsc, T$}{
			Gradient step: Find $\tilde{\vv} _{t}$ from $\label{gradient}
			\nabla \Psi_1 (\tilde{\vv} _{t}) = \nabla \Psi_1 (\vv_{t-1}) - \eta_t \nabla_{\vv = \vv_{t-1}} f(\vv,\uv_{t-1})
			$\\
			Projection step: Compute $\vv'_{t}$ from $\vv'_{t} = \argmin_{\vv \in \Ec_1} B_{\Psi_1} (\vv,\tilde{\vv} _{t})$ \\
			Mixture step: Compute $\vv_{t}$ from $ \vv_{t} = (1 - \kappa)\vv'_{t} + \kappa \vv_0$	
		}
		\label{cbayfvyttyt}
		\caption{ Computation of $\vv_t$.}
		
	\end{algorithm}
	
	\begin{algorithm}[t]
		\SetAlgoLined
		Initialize:  choose $\vv_0,\uv_0$ from \eqref{x0def} and \eqref{y0def}, respectively \\
		\For{$t = 1,2,\dotsc, T$}{
			Gradient step: Find $\tilde{\uv} _{t}$ from $\label{gradient2}
			\nabla \Psi_2 (\tilde{\uv} _{t}) = \nabla \Psi_2 (\uv_{t-1}) - \eta_t \nabla_{\uv = \uv_{t-1}} f(\vv_{t-1},\uv)
			$\\
			Projection step: Compute $\uv'_{t}$ from $\uv'_{t} = \argmin_{\uv \in \Ec_2} B_{\Psi_2} (\uv,\tilde{\uv} _{t})$ \\
			Mixture step: Compute $\uv_{t}$ from $ \uv_{t} = (1 - \kappa)\uv'_{t} + \kappa \uv_0$		
		}
		\label{hprawihte}
		\caption{ Computation of $\uv_t$.}
		
	\end{algorithm}

	Similarly, the range of  $\nabla \Psi_2$ is $\Big\{\uv \in \RR^{J\times K \times K} ~\big|~ u(j,k_1,k_2) = 0 \text{ if } j \notin \Sc_\metric(k_1,k_2) \big\}$, guaranteeing the existence of  $\tilde{\uv} _{t+1}$ in  the gradient step.

	The following result guarantees the convergence of proposed iterative algorithm to the saddlepoint. 
	\begin{proposition}\label{convergence}
		Let $\kappa = \frac{1}{T}$ and the stepsize $\eta_t = \bar \eta= \sqrt{\frac{1}{T}}$. Then, we have
		\begin{align}
			\bigg|f\bigg(\frac{1}{T}\sum_{t=0}^{T-1}\vv_t,\frac{1}{T}\sum_{t=0}^{T-1}&\uv_t \bigg) - \min_{\vv \in \Ec_1} \max_{\uv \in \Ec_2} f(\vv,\uv)\bigg| \notag \\
			&\leq \frac{1}{\sqrt{T}} \Big(4J\log \frac{K}{W_{\rm min}} +G^2\Big).
		\end{align} 
	\end{proposition}
	
	\begin{IEEEproof}
		We assume several properties of Bregman divergences without proof. For further details see \cite{Bubeck}. 
		The first-order optimality of Bregman divergence projections states that for any $\vv_* \in \Ec_1$ \cite{Bubeck}
		\begin{align}\label{bregmannproperty}
			&\langle \gv_t, \vv'_{t+1} - \vv_* \rangle \notag \\
			& \leq \frac{1}{\bar \eta} \Big(B_{\Psi}(\vv_*,\vv_t) - B_{\Psi}(\vv_*,\vv'_{t+1}) - B_{\Psi}(\vv'_{t+1},\vv_t) \Big),
		\end{align}
		where for ease of notation we have defined $\gv_t \eqdef \nabla_{\vv = \vv_t} f(\vv,\uv_t)$. As a result, for any arbitrary $\vv_*\in \Ec_1$ we have
		\begin{align} \label{oghfueasfbe}
			&\sum_{t = 0}^{T-1} \big[f(\vv_t,\uv_t) - f(\vv_*,\uv_t) \big]\notag\\
			&\leq \sum_{t = 0}^{T-1} \langle \gv_t , \vv_t - \vv_* \rangle \\
			&= \sum_{t = 0}^{T-1}\big[\langle \gv_t, \vv_{t+1}' - \vv_* \rangle +  \langle \gv_t, \vv_t - \vv_{t+1}' \rangle \big] \\
			& \leq \sum_{t = 0}^{T-1}\bigg[\frac{1}{\bar \eta} \Big(B_{\Psi_1}(\vv_*,\vv_t) - B_{\Psi_1}(\vv_*,\vv'_{t+1}) - B_{\Psi_1}(\vv'_{t+1},\vv_t) \Big) \notag \\
			&~~~~~+\langle \gv_t, \vv_t - \vv_{t+1}' \rangle\bigg] \\ \label{friuheyrfbqew}
			&\leq \sum_{t = 0}^{T-1}\bigg[\frac{1}{\bar \eta} \Big(B_{\Psi_1}(\vv_*,\vv_t) - B_{\Psi_1}(\vv_*,\vv'_{t+1}) - B_{\Psi_1}(\vv'_{t+1},\vv_t) \Big) \notag\\
			&~~~~~+ \frac{1}{2\bar \eta}|\vv_t - \vv_{t+1}'|_1^2 +  \frac{\bar \eta}{2} G^2\bigg]\\ 
			&\leq \sum_{t = 0}^{T-1}\bigg[\frac{1}{\bar \eta} \Big(B_{\Psi_1}(\vv_*,\vv_t) - B_{\Psi_1}(\vv_*,\vv'_{t+1})\Big) \notag\\
			& ~~~~~ - \frac{1}{2\bar \eta}|\vv'_{t+1}-\vv_t|_1^2  + \frac{1}{2\bar \eta}|\vv_t - \vv_{t+1}'|_1^2 +  \frac{\bar \eta}{2} G^2 \bigg] \\ \label{vowdemwed}
			& =  \sum_{t = 0}^{T-1}\bigg[\frac{1}{\bar \eta} \Big(B_{\Psi_1}(\vv_*,\vv_t) - B_{\Psi_1}(\vv_*,\vv'_{t+1})\Big) +  \frac{\bar \eta}{2} G^2 \bigg],
		\end{align}
		where \eqref{oghfueasfbe} follows from the definition of convexity of $f$, \eqref{friuheyrfbqew} follows from H\"older's inequality \cite{gallager1968information} $\langle \gv_t, \vv_t - \vv_{t+1} \rangle \leq |\vv_t - \vv_{t+1}|_1 |\gv_t|_{\infty} \leq \frac{1}{2\bar \eta}|\vv_t - \vv_{t+1}|_1^2 +  \frac{\bar \eta}{2} |\gv_t|_{\infty}^2$ and $|\gv_t|_{\infty} \leq G$ from Lemma \ref{bounded gradient}. Moreover, inequality \eqref{vowdemwed} follows from \eqref{oxwuewnrnr}.
		Furthermore, from convexity of $B_{\Psi_1}(\cdot,\cdot)$ in the second argument we have that
		\begin{align} \label{coejfief}
			B_{\Psi_1}(\vv_*,\vv_t) &\leq (1-\kappa)B_{\Psi_1}(\vv_*,\vv_t') + \kappa B_{\Psi_1}(\vv_*,\vv_0).
		\end{align}
		Therefore plugging \eqref{coejfief} in \eqref{vowdemwed} we get		
		\begin{align}
			&\sum_{t = 0}^{T-1} \big(f(\vv_t,\uv_t) - f(\vv_*,\uv_t) \big)\notag \\
			&\leq \sum_{t = 0}^{T-1}\bigg[\frac{1}{\bar \eta} \Big((1-\kappa)B_{\Psi_1}(\vv_*,\vv_t') + \kappa B_{\Psi_1}(\vv_*,\vv_0)\notag\\
			&~~~~~~~~~~~~~~~~~~~~~~~~~~~~~- B_{\Psi_1}(\vv_*,\vv'_{t+1})\Big) +  \frac{\bar \eta}{2} G^2 \bigg]  \\
			&= \sum_{t = 0}^{T-1} \frac{1}{\bar \eta} \Big((1-\kappa)B_{\Psi_1}(\vv_*,\vv_t') + \kappa B_{\Psi_1}(\vv_*,\vv_0)\notag\\
			&~~~~~~~~~~~~~~~~~~~~~~~~~~~~~- B_{\Psi_1}(\vv_*,\vv'_{t+1})\Big) + \frac{T\bar \eta}{2}G^2 \\
			& = \sum_{t = 0}^{T-1} \frac{1}{\bar \eta} \Big((1-\kappa)B_{\Psi_1}(\vv_*,\vv_t') - B_{\Psi_1}(\vv_*,\vv'_{t+1})\Big) \notag \\
			&~~~~~~~~~~~~~~~~~~~~~~~~~~~~~+\frac{T \kappa}{\bar \eta} B_{\Psi_1}(\vv_*,\vv_0)+ \frac{T\bar \eta}{2}G^2 \\
			& = \sum_{t = 0}^{T-2} \frac{1}{\bar \eta} \Big( - \kappa B_{\Psi_1}(\vv_*,\vv'_{t+1})\Big) + (1-\kappa)B_{\Psi_1}(\vv_*,\vv_0') \notag \\ 
			&~~~~~~~~~~-B_{\Psi_1}(\vv_*,\vv'_{T}) + \frac{T \kappa}{\bar \eta} B_{\Psi_1}(\vv_*,\vv_0)+ \frac{T\bar \eta}{2}G^2 \\ \label{cvayfyafg}
			& \leq \frac{1}{\bar \eta} B_{\Psi_1}(\vv_*,\vv_0) + \frac{T \kappa}{\bar \eta} B_{\Psi_1}(\vv_*,\vv_0)+ \frac{T\bar \eta}{2}G^2
		\end{align}
		where $\vv'_{0} = \vv_0$ (note that this is consistent with inequality \eqref{coejfief}) and \eqref{cvayfyafg} follows from $ B_{\Psi_1}(\cdot,\cdot)$ being non-negative.
		Therefore, by setting $\kappa =\frac{1}{T} , \bar \eta = \sqrt{\frac{1}{T}}$ and noticing $B_{\Psi_1}(\vv,\vv_0) \leq \log(J) \leq J\log \frac{K}{W_{\rm min}}$ for $J,K>1$ we get 
		\begin{align} \label{vnyrfgbesyrf}
			\frac{1}{T}\sum_{t = 0}^{T-1}\Big( f(\vv_t,\uv_t) &- f(\vv_*,\uv_t)\Big)\notag\\
			& \leq \frac{1}{\sqrt{T}}  \Big(2J\log\frac{K}{W_{\rm min}} + \frac{1}{2}G^2\Big).
		\end{align}
		
		The same procedure gives
		\begin{align} \label{dyrgfeyrrfer}
			\frac{1}{T}\sum_{t = 0}^{T-1}\Big( f(\vv_t,\uv_t) &- f(\vv_t,\uv_*)\Big) \notag\\
			&\geq \frac{-1}{\sqrt{T}}  \Big(2J\log\frac{K}{W_{\rm min}} + \frac{1}{2}G^2\Big).
		\end{align}
		
		As a result, we have
		\begin{align} \label{gujrdtgnuerf4e}
			f\Big(\frac{1}{T}\sum_{t=0}^{T-1} \vv_t,\uv_*\Big) &-  f\Big(\vv_*,\frac{1}{T} \sum_{t=0}^{T-1} \uv_t\Big)\notag\\
			 &\leq  \frac{1}{T}\sum_{t=0}^{T-1} \Big(f(\vv_t,\uv_*) - f(\vv_*,\uv_t)\Big) \\ \label{vmrutgaywer} 
			&\leq \frac{1}{\sqrt{T}} \Big(4J\log\frac{K}{W_{\rm min}} + G^2\Big)
		\end{align}
		where \eqref{gujrdtgnuerf4e} follows from the convex-concave nature of $f$ and \eqref{vmrutgaywer} follows from summing \eqref{vnyrfgbesyrf} and \eqref{dyrgfeyrrfer}.
		
		Since $\vv_*$ and $\uv_*$ were arbitrary, let $\vv_* = \argmin_{\vv \in \Ec_1} f\Big(\vv,\frac{1}{T} \sum_{t=0}^{T-1} \uv_t\Big)$ and $\uv_* = \argmax_{\uv \in \Ec_2} f\Big(\frac{1}{T}\sum_{t=0}^{T-1}\vv_t,\uv\Big)$ then we have
		\begin{align} \label{cfbweufbpg}
			f\Big(\vv_*,\frac{1}{T} \sum_{t=0}^{T-1} \uv_t\Big) &= \min_{\vv \in \Ec_1} f\Big(\vv,\frac{1}{T} \sum_{t=0}^{T-1} \uv_t\Big)\\
			& \leq \min_{\vv \in \Ec_1} \max_{\uv \in \Ec_2} f(\vv,\uv) \\
			& \leq \max_{\uv \in \Ec_2} f\Big(\frac{1}{T}\sum_{t=0}^{T-1}\vv_t,\uv\Big) \\
			&= f\Big(\frac{1}{T}\sum_{t=0}^{T-1}\vv_t,\uv_*\Big).
		\end{align}
		
		In addition, observe that
		\begin{align}\label{efjinvsfr}
			f\Big(\vv_*,\frac{1}{T} \sum_{t=0}^{T-1} \uv_t\Big) &\leq  f\Big(\frac{1}{T}\sum_{t=0}^{T-1}\vv_t,\frac{1}{T} \sum_{t=0}^{T-1} \uv_t\Big)\\
			& \leq f\Big(\frac{1}{T}\sum_{t=0}^{T-1}\vv_t,\uv_*\Big),
		\end{align}
		and therefore by combining \eqref{cfbweufbpg} and \eqref{efjinvsfr} we have
		\begin{align}
			\bigg|f\Big(\frac{1}{T}\sum_{t=0}^{T-1}&\vv_t,\frac{1}{T} \sum_{t=0}^{T-1} \uv_t\Big) -  \min_{\vv \in \Ec_1} \max_{\uv \in \Ec_2} f(\vv,\uv)\bigg| \notag \\
			&\leq f\Big(\frac{1}{T}\sum_{t=0}^{T-1}\vv_t,\uv_*\Big) - f\Big(\vv_*,\frac{1}{T} \sum_{t=0}^{T-1} \uv_t\Big).
		\end{align}
		This combined with \eqref{vmrutgaywer} finishes the proof.
	\end{IEEEproof} 
	
\begin{figure}[htp]
		\centering 
		\input{plot2.tex}
		\caption{Convergence of the proposed iterative algorithm to compute $\bar R_\metric(W)$ for the channel and metric from Example \ref{examp2}.}
		\label{fig:exm} 
	\end{figure}
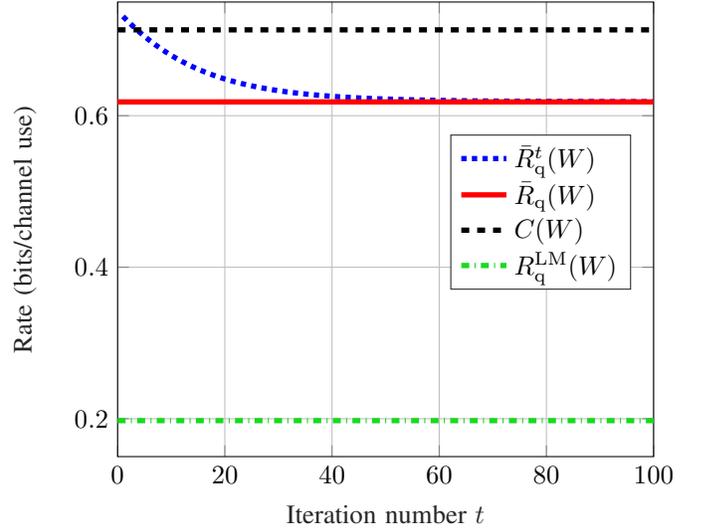

	It is well known that the divergence projection step in Algorithm \ref{cbayfvyttyt} can be computed efficiently as \cite{Bubeck,csiszar2004information}
	\begin{align}
		\label{peojy54u564}
		\vv'_{t+1}(j) = \frac{\tilde{\vv}_{t+1}(j)}{\sum_{j'}\tilde{\vv}_{t+1}(j')}.
	\end{align}
	Similarly, the divergence projection on $ \Mc_{\rm max}(\metric) \cap \{P_{Y\hat{Y}|X}|P_{Y|X} = W\}$ in  Algorithm \ref{hprawihte} can be computed efficiently as
	\begin{align}
		\uv'_{t+1}(j,k_1,k'_2) = \frac{W(k_1|j)\tilde{y}_{t+1}(j,k_1,k_2)\indicator\{j \in \Sc(k_1,k_2)\}}{\sum_{k'_2}\tilde{y}_{t+1}(j,k_1,k'_2)\indicator\{j \in \Sc(k_1,k'_2)\}}.
	\end{align}
	Then the iterative algorithm at iteration $t+1$  is summarized by the following:
	\begin{align}\label{gradientdmoirn34}
		\tilde{\vv}_{t+1} &= \vv_{t} \odot \exp\Big(\frac{1}{\sqrt{T}}\nabla_{\vv = \vv_t} f(\vv,\uv_t)\Big)\\ 
		&\vv'_{t+1}(j) = \frac{\tilde{\vv}_{t+1}(j)}{\sum_{j'}\tilde{\vv}_{t+1}(j')}\\
		&\vv_{t+1} = \frac{T-1}{T}\vv'_{t+1} + \frac{1}{T}\vv_0
	\end{align}
	and
	\begin{align}\label{peojectiofnweu4}
		\tilde{\uv}_{t+1} &= \uv_{t} \odot \exp\Big(\frac{1}{\sqrt{T}}\nabla_{\uv = \uv_t} f(\vv_t,\uv)\Big) \\
		\uv'_{t+1}(j,k_1,k'_2) &= \frac{W(k_1|j)\tilde{\uv}_{t+1}(j,k_1,k_2)\indicator\{j \in \Sc(k_1,k_2)\}}{\sum_{k'_2}\tilde{\uv}_{t+1}(j,k_1,k'_2)\indicator\{j \in \Sc(k_1,k'_2)\}}\\
		&\uv_{t+1} = \frac{T-1}{T}\uv'_{t+1} + \frac{1}{T}\uv_0,
	\end{align}
	where $\av\odot\bv$ denotes componentwise product of the entries of vectors $\av,\bv$. Moreover, \eqref{gradientdmoirn34} and \eqref{peojectiofnweu4}  correspond to the gradient steps.
	Therefore, we can use the gradients computed in the previous section to run the algorithm. 
	%While Theorem guarantees a convergence of order $O\left(\frac{(\log(T))^2}{\sqrt{T}}\right)$, the convergence of the algorithm can be accelerated to order $O\left(\frac{\log(T)}{T}\right)$ with a more careful analysis and choosing different step sizes \cite{Nemirovski}.  
	Figure \ref{fig:exm}  illustrates the convergence of $\bar R_\metric^t(W)$ over the iterations $t$ to the upper bound $\bar R_\metric(W)$, using the suggested iterative algorithm for the channel and metric of Example \ref{examp2}. For reference, we also plot the $C(W)$ and the LM rate $R_\metric^{\rm LM}(W)$. We have chosen an equiprobable maximal joint conditional distribution as initial condition and the step size $\bar \eta = \frac{1}{\sqrt{100}}$.
	
	%%%%%%%%%%%%%%%%%%%%%%%%%%%%%%%%%%%%%%%%%%%

	\section{Multiletter Bound} \label{multilettersec}
	In this section, we study the multiletter extension of the bound \eqref{eq:upper_bound}. In particular, we show that the multiletter version cannot improve on the single-letter bound. We define the $\ell$-letter decoding metric $\metric^{(\ell)}: \Xc^\ell \times \Yc^\ell \to \RR$ as follows,
	\begin{align}
		\metric^{(\ell)}\big((x_1,x_2,\dotsc,x_\ell),(y_1,y_2,\dotsc,y_\ell)\big) = \sum_{i = 1}^{\ell}\metric(x_i,y_i).
	\end{align}
	This decoding metric definition is consistent with the additive decoder we have defined in \eqref{pvigtufunee}. We denote $\jv\in\Xc^\ell$ and $\kv\in\Yc^\ell$ as the $\ell$-letter inputs and outputs, respectively. Let $W^{(\ell)}$ denote a DMC over input alphabet $\Xc^\ell$ and output alphabet $\Yc^\ell$ with the channel rule $W^{(\ell)}\big((y_1,y_2,\dotsc,y_\ell)|(x_1,x_2,\dotsc,x_\ell)\big) = \prod_{i = 1}^{\ell}W(y_i|x_i)$. Additionally, we define $P^{(\ell)}_X$ and $P^{(\ell)}_{Y\hat Y |X}$ accordingly
	\begin{align}
		&P^{(\ell)}_X(x_1,\dotsc x_\ell)= \prod_{i = 1}^{\ell} P_X(x_i)\\
		&P^{(\ell)}_{Y\hat Y |X}\big((y_1,y_2,\dotsc,y_\ell),(\hat{y_1},\hat{y_2},\dotsc,\hat{y_\ell})|(x_1,x_2,\dotsc,x_\ell)\big) \notag \\
		& ~~~~~~~~~~~~~~~~~~~=\prod_{i = 1}^{\ell}P_{Y\hat Y |X}(y_i, \hat{y_i}|x_i) 
	\end{align}
	$X^\ell$ and $Y^\ell, \hat Y^\ell$ denote random variables defined on alphabets $\Xc^\ell$, $\Yc^\ell$ and $\Yc^\ell$, respectively. Moreover, $\Sc_\metric^{(\ell)}(\kv_1,\kv_2)$ is defined as 
	\begin{align}
		&\Sc_\metric^{(\ell)}(\kv_1,\kv_2)\notag\\
		& \eqdef\big \{ \iv \in \Xc^{\ell} \,|\, \iv = \argmax_{\iv'\in\Xc^{\ell}}\metric^{(\ell)}(\iv',\kv_2) - \metric^{(\ell)}(\iv',\kv_1)\big\}.
	\end{align}
	
	In the following lemma we characterize the sets $\Sc_\metric^{(\ell)}(\kv_1,\kv_2)$ and relate them to $\Sc_\metric(k_{1,i},k_{2,i}), i = 1,2,\dotsc,\ell$. 
	\begin{lemma}\label{lemma:sets}
		For $\jv\in\Xc^\ell, \kv_1\in\Yc^\ell, \kv_2\in\Yc^\ell$ we have that $\jv \in \Sc_\metric^{(\ell)}(\kv_1,\kv_2)$ if and only if for all $1 \leq i \leq \ell$ we have
		\begin{align}
			j_i \in \Sc_\metric(k_{1,i},k_{2,i}).
		\end{align}  
	\end{lemma}
	\begin{IEEEproof}
		We have
		\begin{align}
			&\argmax_{\jv \in \Xc^\ell} \metric^{(\ell)}(\jv,\kv_2) - \metric^{(\ell)}(\jv,\kv_1) \notag\\
			&~~~~ = \argmax_{\jv \in \Xc^\ell} \sum_{i = 1}^{\ell}\metric(j_i,k_{2,i}) - \metric(j_i,k_{2,i}) \\ \label{vvotginwefnu}
			&~~~~ = \argmax_{(j_1,j_2,\dotsc,j_\ell) \in \Xc^\ell} \sum_{i = 1}^{\ell}\metric(j_i,k_{2,i}) - \metric(j_i,k_{2,i}).
		\end{align}
		From \eqref{vvotginwefnu} we get that if $(j_1,j_2,\dotsc,j_\ell) \in \Sc_\metric(\kv_1,\kv_2)$ then for all $1 \leq i \leq \ell$ we should have $j_i \in \Sc_\metric(k_{1,i},k_{2,i})$. Therefore, 
		\begin{align}\label{multiple}
			&\Sc_\metric^{(\ell)}(\kv_1,\kv_2) \notag\\
			&= \Sc_\metric(k_{1,1},k_{2,1}) \times \Sc_\metric(k_{1,2},k_{2,2}) \times \cdots \times \Sc_\metric(k_{1,\ell},k_{2,\ell}).
		\end{align}
	\end{IEEEproof}
	
	For the above $\ell$-letter alphabets and distributions, the construction and analysis of the bound remains unchanged. Therefore, \eqref{eq:upper_bound} remains valid for its $\ell$-letter extension, which can be written as
	\begin{align}\label{multiletter}
		\bar R_\metric^{(\ell)}(W) &\triangleq \frac{1}{\ell}\bar R_{\metric^{(\ell)}}(W^{(\ell)})\\ \label{eq:multiletter_long}
		& = \frac{1}{\ell}\max_{P_{X^\ell}}\min_{\substack{P_{Y^\ell\hat{Y}^\ell|X^\ell} \in \Mc_{\rm max}(\metric^{(\ell)})\\ P_{Y^\ell|X^\ell} = W^{(\ell)}}} I(p_{X^\ell},P_{\hat{Y}^\ell|X^\ell})\\ \label{chdsfniadnw}
		& = \frac{1}{\ell} \min_{\substack{P_{Y^\ell\hat{Y}^\ell|X^\ell} \in \Mc_{\rm max}(\metric^{(\ell)})\\ P_{Y^\ell|X^\ell} = W^{(\ell)}}} C(P_{\hat{Y}^\ell|X^\ell}).
	\end{align}
	
	We have the following result.
	\begin{proposition}
		\beq
		\bar R_\metric^{(\ell)}(W) = \bar R_\metric(W).
		\eeq
	\end{proposition}
	
	\begin{IEEEproof}
		Recall that if we find a feasible pair $P_{Y^\ell\hat{Y}^\ell|X^\ell},P_{X^\ell}$ such that when fixing $P_{Y^\ell\hat{Y}^\ell|X^\ell}$, the input distribution $P_{X^\ell}$ is a maximizer of $f(\cdot,P_{Y^\ell\hat{Y}^\ell|X^\ell})$,  and when fixing $P_{X^\ell}$, the joint conditional distribution $P_{Y^\ell\hat{Y}^\ell|X^\ell}$ is a minimizer of $f(p_{X^\ell},\cdot)$, then the pair $(P_{X^\ell},P_{Y^\ell\hat{Y}^\ell|X^\ell})$ is a saddlepoint. Therefore, we can use the mentioned property to show $ P^{*(\ell)}_X, P_{Y\hat Y|X}^{*(\ell)}$ is a saddlepoint for the multiletter bound.
		As a result, it is sufficient to show that  $P_{Y\hat Y|X}^{*(\ell)}$ is a minimizer of \eqref{eq:multiletter_long} by fixing $ P^{*(\ell)}_X$ and additionally, $P^{*(\ell)}_X$ is the maximizer of \eqref{eq:multiletter_long} by fixing $P_{Y\hat Y|X}^{*(\ell)}$. 
		
		Firstly, we verify the claim that  $P^{*(\ell)}_X$ is the maximizer of \eqref{eq:multiletter_long} by fixing $P_{Y\hat Y|X}^{*(\ell)}$.
		The validity of this claim follows from \eqref{chdsfniadnw} and the fact that $\frac{1}{\ell}C( P_{\hat Y|X}^{*(\ell)})=C(P^*_{\hat Y|X})$ with the product distribution $P^{*(\ell)}_X$ being the capacity-achieving distribution of $C( P_{\hat Y|X}^{*(\ell)})$. 
		
		Secondly, we verify the claim that  $P_{Y\hat Y|X}^{*(\ell)}$ is a minimizer of \eqref{eq:multiletter_long} by fixing $ P^{*(\ell)}_X$. This is shown in the following lemma. We prove that by fixing $ P^{*(\ell)}_X$, then $ P_{Y\hat Y|X}^{*(\ell)}$  satisfies the KKT conditions and hence, it is a minimizer of \eqref{eq:multiletter_long}. Before stating the result, we recall that the multiletter counterparts of the single-letter KKT conditions given in \eqref{firstoptimal} and \eqref{secondoptimal} hold. Moreover, as in the single-letter case, the multiletter KKT conditions are sufficient for global optimality, because the function $f( P^{*\ell}_X,\cdot)$ is concave and optimization constraints are affine \cite{boyd}. Using Lemma \ref{lemma:multiletter} below completes the proof.
	\end{IEEEproof}

	\begin{lemma}
		\label{lemma:multiletter}
		Let $ P_X^*, P_{Y\hat{Y}|X}^*$ be a saddlepoint for optimization problem \eqref{eq:upper_bound}. Set $P_{X^{\ell}} = P^{*(\ell)}_X$. Then, the joint conditional distribution $ P_{Y\hat{Y}|X}^{*(\ell)}$ is a minimizer of
		\begin{align}
			\min_{\substack{P_{Y^\ell\hat{Y}^\ell|X^\ell} \in \Mc_{\rm max}(\metric^{(\ell)})\\ P_{Y^\ell|X^\ell} = W^{(\ell)}}} I\big(P^{*(\ell)}_X,P_{Y^\ell\hat{Y}^\ell|X^\ell}\big).
			\label{eq:min_multiletter}
		\end{align}
	\end{lemma}
	\begin{IEEEproof}
		We should show that by setting $P_{X^\ell} = P^{*(\ell)}_X$, the multiletter versions of the KKT conditions \eqref{firstoptimal} and \eqref{secondoptimal} hold for $ P_{Y\hat{Y}|X}^{*(\ell)}$. Generalizing the conditions of \eqref{firstoptimal} and \eqref{secondoptimal} to the multiletter case, and setting $P_{Y^\ell\hat Y^\ell|X^\ell} = P^{*(\ell)}_{Y\hat Y|X}$, we should show that for all $\jv,\kv_1 \in \Xc^\ell \times \Yc^\ell$ there exist multipliers  $\lambda_{\jv,\kv_1}$ such that the conditions below are fulfilled. If we show this, then the lemma is proved because these are precisely the conditions for the minimizer of \eqref{eq:min_multiletter}.
		\begin{enumerate}
			\item When $ P_{Y\hat{Y}|X}^{*(\ell)}(\kv_1,\kv_2|\jv) > 0$ we must have,
			\begin{align}\label{tvpdmcergbrf}
				\hspace{-3mm}\frac{\partial}{\partial P_{Y^\ell\hat{Y}^\ell|X^\ell}(\kv_1,\kv_2|\jv)}&f( P_X^{*(\ell)},P_{Y^\ell\hat{Y}^\ell|X^\ell})\bigg|_{P_{Y^\ell\hat Y^\ell|X^\ell} = P^{*(\ell)}_{Y\hat Y|X}}\notag\\
				& = \lambda_{\jv,\kv_1}.
			\end{align}
			\item When $ P_{Y\hat{Y}|X}^{*(\ell)}(\kv_1,\kv_2|\jv) = 0$ and $\jv \in \Sc_\metric^{(\ell)}(\kv_1,\kv_2)$ we must have that,
			\begin{align}\label{vptmgirng}
				\hspace{-3mm}\frac{\partial}{\partial P_{Y^\ell\hat{Y}^\ell|X^\ell}(\kv_1,\kv_2|\jv)}&f( P_X^{*(\ell)},P_{Y^\ell\hat{Y}^\ell|X^\ell})\bigg|_{P_{Y^\ell\hat Y^\ell|X^\ell} =  P^{*(\ell)}_{Y\hat Y|X}}\notag\\
				& \geq \lambda_{\jv,\kv_1}.
			\end{align}
		\end{enumerate}

		Similarly to \eqref{eq:partial_single}, the derivative in \eqref{tvpdmcergbrf} and \eqref{vptmgirng} is given by,
		\begin{align}
			\frac{\partial}{\partial P_{Y^\ell\hat{Y}^\ell|X^\ell}(\kv_1,\kv_2|\jv)}&f( P_X^{*(\ell)}, P_{Y^\ell\hat{Y}^\ell|X^\ell})\bigg|_{P_{Y^\ell\hat Y^\ell|X^\ell} =  P^{*(\ell)}_{Y\hat Y|X}} \notag\\
			&=  P^{*(\ell)}_X(\jv) \log \frac{ P^{*(\ell)}_{\hat Y|X}(\kv_1|\jv)}{ Q^{*(\ell)}_{\hat Y}(\kv_1)}
		\end{align}
		which, by using that in $P_{Y^\ell\hat{Y}^\ell|X^\ell} = P_{Y\hat{Y}|X}^{*(\ell)}$ and $ Q^{*(\ell)}_{\hat Y}$ are product distributions, gives,
		\begin{align}
			&P^{*(\ell)}_X(\jv) \log \frac{ P^{*(\ell)}_{\hat Y|X}(\kv_1|\jv)}{ Q^{*(\ell)}_{\hat Y}(\kv_1)}\notag\\
			&=  P_X^*(j_1) P_X^*(j_2)\cdots P_X^*(j_\ell) \cdot\left( \sum_{i=1}^{\ell} \log \frac{P_{\hat Y|X}^*(k_{2,i}|j_i)}{Q_{\hat Y}^*(k_{2,i})}\right).
			\label{eq:derivative_mult}
		\end{align}

		In order to show that there exist some coefficients $\lambda_{\jv,\kv_1} $ satisfying both \eqref{tvpdmcergbrf} and \eqref{vptmgirng}, we make a specific choice and show that this choice satisfies both \eqref{tvpdmcergbrf} and \eqref{vptmgirng}. To this end, define
		\begin{align}
			\lambda_{\jv,\kv_1} = \begin{cases} 0 \ \ & \prod_{i=1}^{\ell} P_X^*(j_i)= 0 \\
				\prod_{i=1}^{\ell}P_X^*(j_i) \left( \sum_{i=1}^{\ell}\frac{\lambda_{j_i,k_{1,i}}}{ P_X^*(j_i)}\right) \ \ & \prod_{i=1}^{\ell} P_X^*(j_i)\neq 0
			\end{cases} 
		\end{align} 
		where $\lambda_{j_i,k_{1,i}}$ is the single-letter Lagrange multiplier corresponding to $j_i$ and $k_{1,i}$.

		Excluding the cases where $ P_X^*(j_1) P_X^*(j_2)\cdots P_X^*(j_\ell) = 0$ that from \eqref{eq:derivative_mult}, \eqref{tvpdmcergbrf} and \eqref{vptmgirng} the KKT conditions clearly hold, we have two cases
		\begin{enumerate}
			\item When $P^{*(\ell)}_{Y\hat{Y}|X}(\jv,\kv_1,\kv_2) > 0$, then for all $1 \leq i \leq \ell$ we must have $ P_{Y\hat{Y}|X}^*(k_{1,i},k_{2,i}|j_i) > 0$ and therefore, \eqref{firstoptimal} is valid. We have to verify that this implies that \eqref{tvpdmcergbrf} is also valid. As a result,
			\begin{align} 
				&\hspace{-3mm}\frac{\partial}{\partial P_{Y^\ell\hat{Y}^\ell|X^\ell}(\kv_1,\kv_2|\jv)}f( P_X^{*(\ell)},P_{Y^\ell\hat{Y}^\ell|X^\ell})\bigg|_{P_{Y^\ell\hat Y^\ell|X^\ell} =  P^{*(\ell)}_{Y\hat Y|X}} \notag\\
				&~~~ = 	\prod_{i=1}^{\ell}P_X^*(j_i) \left( \sum_{i=1}^{\ell} \log \frac{P_{\hat Y|X}^*(k_{2,i}|j_i)}{Q_{\hat Y}^*(k_{2,i})}\right)\\ \label{vpefmwiefnwre}
				&~~~= 	\prod_{i=1}^{\ell}P_X^*(j_i) \left( \sum_{i=1}^{\ell}\frac{\lambda_{j_i,k_{1,i}}}{ P_X^*(j_i)}\right) \\
				&~~~= \lambda_{\jv,\kv_1} 
			\end{align}
			where \eqref{vpefmwiefnwre} holds from the single-letter optimality in \eqref{firstoptimal}.
			\item When $ P_{Y\hat{Y}|X}^{*(\ell)}(\kv_1,\kv_2|\jv) = 0$ and $\jv \in \Sc_\metric^{(\ell)}(\kv_1,\kv_2)$, as a result of the Lemma \ref{lemma:sets}, we have that $\Sc_\metric^{(\ell)}(\kv_1,\kv_2)$ is a product set, \textit{i.e.} for all $1 \leq i \leq \ell$,
			\begin{align}
				j_i \in \Sc_\metric(k_{1,i},k_{2,i}).
			\end{align}
			Moreover, either $P_{Y\hat{Y}|X}^*(k_{1,i},k_{2,i}|j_i) > 0$ where \eqref{firstoptimal} is satisfied or $ P_{Y\hat{Y}|X}^*(k_{1,i},k_{2,i}|j_i) = 0$ where \eqref{secondoptimal} is satisfied.
			As a result, with these assumptions in mind, we should verify that \eqref{vptmgirng} is valid. We have
			\begin{align} 
				&\hspace{-3mm}\frac{\partial}{\partial P_{Y^\ell\hat{Y}^\ell|X^\ell}(\kv_1,\kv_2|\jv)}f(P_X^{*(\ell)},P_{Y^\ell\hat{Y}^\ell|X^\ell})\bigg|_{P_{Y^\ell\hat Y^\ell|X^\ell} = P^{*(\ell)}_{Y\hat Y|X}} \notag\\
				&~~~ = \prod_{i=1}^{\ell}P_X^*(j_i) \left( \sum_{i=1}^{\ell} \log \frac{P_{\hat Y|X}^*(k_{2,i}|j_i)}{Q_{\hat Y}^*(k_{2,i})}\right)\\ \label{itmgurwufuw}
				&~~~\geq \prod_{i=1}^{\ell}P_X^*(j_i) \left( \sum_{i=1}^{\ell}\frac{\lambda_{j_i,k_{1,i}}}{ P_X^*(j_i)}\right) \\
				&~~~= \lambda_{\jv,\kv_1} 
			\end{align}
			where \eqref{itmgurwufuw} is true because of the single-letter optimality in \eqref{firstoptimal} and \eqref{secondoptimal}.
			
		\end{enumerate}
	\end{IEEEproof}

	\section*{Appendix A}
	In this appendix we  provide the proof of Theorem \ref{odjehrf7erqwe}.
	Without loss of generality, we assume that the sequence $\big\{\metric(1,k) - \metric(2,k)\big\}_{k=1}^{K}$ is non-decreasing, \textit{i.e.} for $k_1 \leq k_2$,
	\begin{align}\label{frtgeth}
		\metric(1,k_1) - \metric(2,k_1) \leq \metric(1,k_2) - \metric(2,k_2).
	\end{align}
	We can assume this, since it is always possible to relabel the output alphabet such that this property is fulfilled.
	This assumption simplifies the  evaluation of the sets $\Sc(\cdot,\cdot)$.
	For $k_1 = k_2$ we have $\Sc(k_1,k_2) = \{1,2\}$.
	Moreover, when $k_1 < k_2$ from \eqref{frtgeth} and Definition \ref{fewer4tqw}, we have that $1 \in \Sc(k_1,k_2)$ and $2 \in \Sc(k_2,k_1)$.

	We prove a slightly stronger result. In particular, we prove that the condition $C_\metric(W) = C(W)$ implies that sequences 
	\begin{align}
		\Big\{ P_X^\star(1)\log\frac{W(k|1)}{\hat Q_{\hat Y}(k)}\Big\}_{k=1}^{K}, ~\Big\{- P_X^\star(2)\log\frac{W(k|2)}{ Q^\star_{\hat Y}(k)}\Big\}_{k=1}^{K}
	\end{align}
	both should have the  same order as the decoding metric difference sequence $\{\metric(1,k) - \metric(2,k)\}_{k=1}^{K}$, where the notation $P_X^\star$ refers to the capacity-achieving distribution of $W$; $Q^\star$ denotes the corresponding output distribution.
	
	Assume that $C_\metric(W) = C(W)$. Therefore, $P_X^\star, P_{Y\hat Y|X}=P_{YY|X}$ must be a saddlepoint of \eqref{ferf3r3}. As a result, the KKT conditions in \eqref{firstoptimal} and \eqref{secondoptimal} must hold. Observe  that
	\begin{align}
		P_{YY|X}(k_1,k_2|j) = \begin{cases} W(k_1|j) \ &k_1 = k_2\\
			0 &k_1 \neq k_2.
		\end{cases}
	\end{align}
	Therefore, combining the KKT conditions in \eqref{firstoptimal} and \eqref{secondoptimal} we obtain,
	\begin{enumerate}
		\item If $k_1 = k_2$, for both $j = 1,2$  we have 
		\begin{align}
			P_X^\star(j) \log \frac{W(k_1|j)}{\hat Q_{Y}(k_1)} = \lambda_{j,k_1}
		\end{align}
		\item If $k_1 < k_2$ we know $1 \in \Sc(k_1,k_2)$ and $2 \in \Sc(k_2,k_1)$, therefore,
		\begin{align}
			& P_X^\star(1) \log \frac{W(k_2|1)}{ Q^\star_{Y}(k_2)} \geq \lambda_{1,k_1}\\
			& P_X^\star(2) \log \frac{W(k_1|2)}{ Q^\star_{Y}(k_1)} \geq \lambda_{2,k_2}.
		\end{align} 
	\end{enumerate}
	As a result, if $k_1 < k_2$ 
	\begin{align}
		P_X^\star(1) \log \frac{W(k_2|1)}{ Q^\star_{Y}(k_2)} \geq \lambda_{1,k_1} =  P^\star_X(1) \log \frac{W(k_1|1)}{ Q^\star_{Y}(k_1)} 
	\end{align}
	\begin{align}
		P_X^\star(2) \log \frac{W(k_1|2)}{ Q^\star_{Y}(k_1)} \geq \lambda_{2,k_2} =  P^\star_X(2) \log \frac{W(k_2|2)}{ Q^\star_{Y}(k_2)}.
	\end{align}
	Therefore, we get that  $\Big \{ P^\star_X(1)\log\frac{W(k|1)}{ Q^\star_{Y}(k)}\Big \}_{k=1}^{K}$ and $-\Big \{ P^\star_X(2)\log\frac{W(k|2)}{ Q^\star_{Y}(k)}\Big \}_{k=1}^{K}$ are both non-decreasing sequences and so is any linear combination of them with positive coefficients. Therefore, since
	\begin{align}
		\log W(k|1) - \log W(k|2) &= \frac{1}{ P^\star_X(1)} \Big( P^\star_X(1)\log\frac{W(k|1)}{ Q^\star_{ Y}(k)}
		\Big) \nonumber\\
		&- \frac{1}{ P^\star_X(2)} \Big( P^\star_X(2)\log\frac{W(k|2)}{ Q^\star_{ Y}(k)}\Big)
	\end{align}
	we conclude that the sequence $\{\log W(k|1) - \log W(k|2)\}_{k=1}^{K}$ is a non-decreasing sequence.
	
	%%%%%%%%%%%%%%%%%%%%%%%%%%%%%%%%
	
	\section*{Appendix B}
	This section is addresses the choice of $\epsilon$ in proof of the main theorem in Section \ref{sec:end}.
	Let $f:\Ac \to \RR$ be a continuous function and $\Ac$ be a compact set. Then this function is uniformly continuous.  We apply this fact to entropy function $H:\Delta^J \to \RR$ where $\Delta^J = \{\xv \in \RR^J |x_i \geq 0, i=1,2,\dotsc, J \}$ is the $J$-dimensional simplex. Therefore, for any $\delta>0$ there exists an $\epsilon > 0$ such that for any $\pv_1,\pv_2 \in \Delta^J$  that are $\epsilon$-close {i.e.} $|\pv_1 - \pv_2|_{\infty} \leq \epsilon$ we have 
	\begin{align}
		\big|H(\pv_1)-H(\pv_2)\big| \leq\delta.
	\end{align}
	Let $\Vm$ be matrix of a conditional distribution and $\Vm_1,\Vm_2,\dotsc,\Vm_J$ be rows of $\Vm$. Consider any type $\pv$, any conditional distribution matrix $\hat \Vm$ with rows $\hat \Vm_1,\hat \Vm_2,\dotsc,\hat \Vm_J$ and let $\qv, \hat \qv$ be output distributions corresponding to input type $\pv$ and channels $\Vm,\hat{\Vm}$, respectively. Then, we have
	\begin{align}
		\big|\qv-\hat{\qv}\big|_{\infty} &\leq \big|\Vm - \hat{\Vm}\big|_{\infty}\\
		\big|\Vm_i - \hat \Vm_i\big|_{\infty} &\leq |\Vm - \hat \Vm\big|_{\infty}.
	\end{align}
	As a result, if $\big|\Vm - \hat{\Vm}\big|_{\infty} \leq \epsilon$ we get
	\begin{align}
		\big|H(\qv)-H(\hat{\qv})\big|_{\infty} &\leq \delta \\
		\big|H(\Vm_i) - H(\hat \Vm_i)\big| &\leq \delta.
	\end{align}
	As for $H(\Vm|\pv)$ we have
	\begin{align}
		H(\Vm|\pv) = \sum_{j=1}^{J}\pv(j)H(\Vm_j),
	\end{align}
	and thus,
	\begin{align}
		\big|H(\Vm|\pv)-H(\hat{\Vm}|\pv)\big|_{\infty} &\leq \sum_{j=1}^{J}\pv(i)\big|H(\Vm_j)-H(\hat{\Vm}_j)\big|_{\infty}\\
		& \leq \delta.
	\end{align}
	Setting $\delta =  \frac{\sigma}{4}$ gives the result.
	
		%%%%%%%%%%%%%%%%%%%%%%%%%%%%%%%%
		\section*{Appendix C}
	In this appendix we discuss the case where some entries of the decoding metric matrix are $-\infty$. When computing the set $\Sc_{\metric}(k_1,k_2)$ we compare expressions that contain $-\infty$ using the following rules:
	\begin{enumerate}
\item $-\infty - (-\infty) = -\infty - (-\infty)$ is a tie \label{item:prop1}
\item $a - (-\infty) > -\infty - (-\infty)$ \label{item:prop2}
\item $-\infty - (-\infty) > -\infty - a$ \label{item:prop3}
\item $a - (-\infty) > b$ and $-\infty -a<b$\label{item:prop4}
\item $-\infty - a < -\infty - b$ if $a> b$, and $-\infty - a = -\infty - b$ if $a=b$ \label{item:prop5}
\item $a-(-\infty) < b -(-\infty)$ if $a<b$, and $a-(-\infty) = b -(-\infty)$ if $a=b$ \label{item:prop6}
\item $0\cdot (-\infty) = 0$ \label{item:prop7}
\end{enumerate}
where $a,b\in\RR$.

%	 we consider $-\infty - (-\infty)$ being a tie to another $-\infty - (-\infty)$, less than $a - (-\infty)$, larger than $-\infty - a$, when $a$ is finite and less than any other finite value. Also we consider $a - (-\infty)$ being larger than any finite value. Additionally, $-\infty - a \leq -\infty - b$ if $b \leq a$ for $a,b \in \RR$. We have summerized the rules as follows. for any $a,b \in \RR$
%	 \begin{align} \label{naefoajfnad}
%	 	-\infty - (-\infty) &= -\infty - (-\infty) \\
%	 	-\infty - (-\infty) &< a - (-\infty) \\
%	 	-\infty - (-\infty) &> -\infty - a \\
%	 	a - (-\infty) &> b \\ \label{adokaocjeif}
%	 	-\infty - a &\leq -\infty - b \text{ if } b \leq a
%	 \end{align}

As we show next, Lemma  \ref{lemmaximal} remains true for this case.	Observe that in the decomposition 
 	\begin{align}
 		\metric(\xv,\yv) = \sum_{j,k} \hat \pv_{\xv,\yv}(j,k) \metric(j,k)
 	\end{align}
    $0\cdot(-\infty)=0$ according to rule \ref{item:prop7}).
	With assumptions of Lemma \ref{lemmaximal} we have that, 
	\begin{align}
		&\metric(\hat{\xv},\hat{\yv}) - \metric(\hat{\xv},\yv) \notag\\
		&= n\sum_{j,k_1,k_2}\hat{\pv}_{\hat{\xv}\yv\hat{\yv}}(j,k_1,k_2)\big(\metric(j,k_2) - \metric(j,k_1)\big)\label{werqtC}\\ \label{ld;ejjreirC}
		&\leq n\sum_{k_1,k_2}\Big(\sum_{j}\hat{\pv}_{\hat{\xv}\yv\hat{\yv}}(j,k_1,k_2)\Big)\max_{j'}\big(\metric(j',k_2) - \metric(j',k_1)\big) \\ \label{sa;efqC}
		&= n\sum_{k_1,k_2}\Big(\sum_{j}\hat{\pv}_{\xv\yv\hat{\yv}}(j,k_1,k_2)\Big)\max_{j'}\big(\metric(j',k_2) - \metric(j',k_1)\big)  \\ \label{fwefneC}
		&= n\sum_{k_1,k_2}\sum_{j}\hat{\pv}_{\xv\yv\hat{\yv}}(j,k_1,k_2)\big(\metric(j,k_2) - \metric(j,k_1)\big) \\ \label{fr4werqC}
		&= \metric(\xv,\hat{\yv}) - \metric(\xv,\yv)
	\end{align}
where in \eqref{werqtC} when upperbounding $\metric(j,k_2) - \metric(j,k_1)$ with $\max_{j'}\big(\metric(j',k_2) - \metric(j',k_1)\big)$ if neither of $\metric(j,k_2), \metric(j,k_1)$ is equal to $-\infty$ the argument remains valid. Moreover,
	 rules \ref{item:prop1}),  \ref{item:prop5}) and \ref{item:prop6}) imply that if $\metric(j,k_1) = -\infty$ in \eqref{werqtC} then $\metric(j',k_1) = -\infty$ for maximizing $j'$. Therefore, if
	$\metric(\hat{\xv},\yv) = -\infty$ then  $\metric(\xv,\yv) = -\infty$. Finally, in \eqref{werqtC} if $\metric(j,k_2) = -\infty$ and $\metric(j,k_1)$ is finite, then $\metric(j',k_1)$ in \eqref{ld;ejjreirC} for maximizing $j'$ is also finite and $\metric(j',k_1) \leq \metric(j,k_1)$. As a result, $ \metric(\xv,\yv) \leq \metric(\hat{\xv},\yv)$.
	
	%%%%%%
	%% Appendix:
	%% If needed a single appendix is created by
	%%
	%\appendix
	%%
	%% If several appendices are needed, then the command
	%%
	% \appendices
	%%
	%% in combination with further \section-commands can be used.
	%%%%%%

	%\section*{Acknowledgment}
	
	%Thanks to ...

	%%%%%%
	%% To balance the columns at the last page of the paper use this
	%% command:
	%%
	%\enlargethispage{-1.2cm} 
	%%
	%% If the balancing should occur in the middle of the references, use
	%% the following trigger:
	%%
	
	%% which triggers a \newpage (i.e., new column) just before the given
	%% reference number. Note that you need to adapt this if you modify
	%% the paper.  The "triggered" command can be changed if desired:
	%%
	%\IEEEtriggercmd{\enlargethispage{-20cm}}
	%%
	%%%%%%

	%%%%%%
	%% References:
	%% We recommend the usage of BibTeX:
	%%
	\bibliographystyle{IEEEtran}
	\bibliography{bibliography}

	%\bibliography{definitions,bibliofile}
	%%
	%% where we here have assume the existence of the files
	%% definitions.bib and bibliofile.bib.
	%% BibTeX documentation can be obtained at:
	%% http://www.ctan.org/tex-archive/biblio/bibtex/contrib/doc/
	%%%%%%
	%% Or you use manual references (pay attention to consistency and the
	%% formatting style!):

\end{document}

%% file: plot2.tex
\definecolor{mycolor1}{rgb}{0.00000,0.44700,0.74100}%
\definecolor{mycolor2}{rgb}{0.85000,0.32500,0.09800}%
\definecolor{mycolor3}{rgb}{0.92900,0.69400,0.12500}%
\definecolor{mycolor4}{rgb}{0.49400,0.18400,0.55600}%

\definecolor{mygreen}{rgb}{0.10000,0.8500,0.12100}%

\begin{tikzpicture}

\begin{axis}[%
width=2.80521in,
height=2.381in,
at={(0.758in,0.666in)},
scale only axis,
xmin=0,
xmax=100,
xlabel style={font=\color{white!15!black}},
xlabel={Iteration number $t$},
ymin=0.15,
ymax=0.75,
ylabel style={font=\color{white!15!black}},
ylabel={Rate (bits/channel use)},
axis background/.style={fill=white},
%axis x line*=bottom,
%axis y line*=left,
xmajorgrids,
xminorgrids,
ymajorgrids,
yminorgrids,
legend style={at={(0.6207,0.3672)}, anchor=south west, legend cell align=left, align=left, draw=white!15!black}
]

axis background/.style={fill=white},
xmajorgrids,
ymajorgrids,

\addplot [color=blue, dotted, line width=2.0pt]
  table[row sep=crcr]{%
1 0.731025481147065\\
2 0.723928444453685\\
3 0.717196762196894\\
4 0.710826257061898\\
5 0.704809585384371\\
6 0.699137011753972\\
7 0.693797048258959\\
8 0.688776976303968\\
9 0.684063268101524\\
10 0.679641923568749\\
11 0.675498736702854\\
12 0.671619503748265\\
13 0.667990183734046\\
14 0.664597020335141\\
15 0.661426632541929\\
16 0.658466080330053\\
17 0.655702910409326\\
18 0.653125186187965\\
19 0.650721505301308\\
20 0.648481007404187\\
21 0.646393374394366\\
22 0.644448824802538\\
23 0.642638103735601\\
24 0.640952469479606\\
25 0.639383677644062\\
26 0.637923963549638\\
27 0.636566023417691\\
28 0.63530299480536\\
29 0.634128436638178\\
30 0.633036309118636\\
31 0.632020953730112\\
32 0.631077073508007\\
33 0.630199713711557\\
34 0.629384242998674\\
35 0.628626335180878\\
36 0.627921951614839\\
37 0.62726732427021\\
38 0.626658939499761\\
39 0.626093522526595\\
40 0.625568022654135\\
41 0.625079599197125\\
42 0.62462560812593\\
43 0.624203589411593\\
44 0.623811255055321\\
45 0.623446477783076\\
46 0.623107280383704\\
47 0.622791825667342\\
48 0.622498407019669\\
49 0.622225439526808\\
50 0.621971451645265\\
51 0.62173507739119\\
52 0.621515049023328\\
53 0.62131019019439\\
54 0.62111940954601\\
55 0.620941694723086\\
56 0.62077610678399\\
57 0.620621774983918\\
58 0.620477891909505\\
59 0.620343708943679\\
60 0.620218532040657\\
61 0.620101717791889\\
62 0.619992669764677\\
63 0.619890835096111\\
64 0.619795701325864\\
65 0.619706793452274\\
66 0.619623671196992\\
67 0.619545926464314\\
68 0.619473180982131\\
69 0.619405084112182\\
70 0.619341310818061\\
71 0.619281559780127\\
72 0.619225551647135\\
73 0.619173027415084\\
74 0.619123746924337\\
75 0.619077487466699\\
76 0.619034042494644\\
77 0.618993220425429\\
78 0.618954843533313\\
79 0.618918746923529\\
80 0.618884777582147\\
81 0.6188527934963\\
82 0.618822662839692\\
83 0.618794263218607\\
84 0.618767480974007\\
85 0.61874221053561\\
86 0.618718353824111\\
87 0.618695819698017\\
88 0.618674523441786\\
89 0.618654386292212\\
90 0.61863533500023\\
91 0.618617301425491\\
92 0.618600222161279\\
93 0.618584038187501\\
94 0.61856869454965\\
95 0.618554140061801\\
96 0.618540327031843\\
97 0.618527211007263\\
98 0.618514750539961\\
99 0.618502906968646\\
100 0.618491644217497\\};
\addlegendentry{$\bar{R}^t_\metric(W)$}

\addplot [color=red, line width=2.0pt]
  table[row sep=crcr]{%
0	0.6182\\
100	0.6182\\
};
\addlegendentry{$\bar R_\metric(W)$}

\addplot [color=black, dashed, line width=2.0pt]
  table[row sep=crcr]{%
0	0.7133\\
100	0.7133\\
};
\addlegendentry{$C(W)$}

\addplot [color=mygreen, dashdotted, line width=2.0pt]
  table[row sep=crcr]{%
0	0.1975\\
100	0.1975\\
};
\addlegendentry{$R_\metric^{\rm LM}(W)$}

\end{axis}

%\begin{axis}[%
%width=5.833in,
%height=4.375in,
%at={(0in,0in)},
%scale only axis,
%xmin=0,
%xmax=1,
%ymin=0,
%ymax=1,
%axis line style={draw=none},
%ticks=none,
%axis x line*=bottom,
%axis y line*=left,
%legend style={legend cell align=left, align=left, draw=white!15!black}
%]
%!TEX encoding = UTF-8 Unicode\end{axis}
\end{tikzpicture}%